\documentclass{nature}

\usepackage{amsmath,amssymb,graphicx,psfrag}

\newcommand{\Pe}{\textrm{Pe}}

\newcommand{\suba}{{a}}
\newcommand{\subb}{{b}}
\newcommand{\subc}{{c}}
\newcommand{\subd}{{d}}
\newcommand{\sube}{{e}}
\newcommand{\subf}{{f}}

\newcommand{\subh}{{h}}

\newcommand{\utv}{\textmu m/s}
\newcommand{\utmicron}{\textmu m}
\newcommand{\utmm}{mm}
\newcommand{\utG}{K/cm}
\newcommand{\uth}{h}

\title{Cell invasion during competitive growth of polycrystalline solidification patterns}

\author{Younggil Song$^{1,2}$, Fatima L. Mota$^3$, Damien Tourret$^4$, Kaihua Ji$^1$, Bernard Billia$^3$, Rohit Trivedi$^5$, Nathalie Bergeon$^3$, \& Alain Karma$^{1,*}$  }

\begin{document}
\maketitle

\begin{affiliations}
 \item Department of Physics and Center for Interdisciplinary Research on Complex Systems, Northeastern University, Boston, MA, USA
 \item Materials Science Division, Lawrence Livermore National Laboratory, Livermore, CA 94550, USA
 \item Aix-Marseille Univ., Université de Toulon, CNRS, IM2NP, Marseille, France
 \item IMDEA Materials Institute, Getafe, Madrid, Spain
 \item Department of Material Science and Engineering, Iowa State University, Ames, IA, USA

$^*$ e-mail: a.karma@northeastern.edu
\newline
\end{affiliations}

\section*{Abstract} 

\begin{abstract}
Spatially extended cellular and dendritic array structures forming during solidification processes such as casting, welding, or additive manufacturing are generally polycrystalline. Both the array structure within each grain and the larger scale grain structure determine the performance of many structural alloys. How those two structures coevolve during solidification remains poorly understood.
By {{{in situ}}} observations of microgravity alloy solidification experiments onboard the International Space Station, we have discovered that individual cells from one grain can unexpectedly invade a nearby grain of different misorientation, either as a solitary cell or as rows of cells. This invasion process causes grains to interpenetrate each other and hence grain boundaries to adopt highly convoluted shapes. Those observations are reproduced by phase-field simulations further demonstrating that invasion occurs for a wide range of misorientations. Those results fundamentally change the traditional conceptualization of grains as distinct regions embedded in three-dimensional space. 
\end{abstract}



\section*{Introduction}

Boundaries emerge in a wide range of pattern forming systems~\cite{CrossHohenberg93}, e.g. during Rayleigh-B\'enard convection~\cite{GreensideEtAl82}, biological tissue development~\cite{Turing52, DahmannEtAl11, Anirban22}, tumor growth with cells migrating in foreign tissues~\cite{FriedlWolf03,CondeelisPollard06}, self-assembly during the growth of colloidal crystals~\cite{XieLiu09, Vogel15}, polycrystalline growth in polymeric and complex fluids~\cite{GranasyEtAl04}, or solidification processing of metallic alloys~\cite{Flemings74, KurzFisher, DantzigRappaz09, MartinEtAl17, pham2020}. 
The formation and evolution of boundaries in pattern-forming systems result from the different properties and mobilities of the patterns constituting the boundary.
New regions may appear during the migration of a boundary, as in the case of freckles in metal casting~\cite{PollockTin06} or that of invasion of cancerous tumors~\cite{FriedlWolf03,CondeelisPollard06}.
The formation and morphological evolution of these boundaries are in turn crucial for the behavior of biological systems~\cite{Turing52,DahmannEtAl11, Anirban22} and the properties of structural materials~\cite{Flemings74,Hirth72,PollockTin06,pham2020}. 

In solidification processing, such as casting, welding, or additive manufacturing, cellular or dendritic patterns develop during the growth of a solid crystal from the liquid phase~\cite{DantzigRappaz09,GranasyEtAl04,AkamatsuNguyen16,pham2020}. 
In metallurgy, each pattern region of similar crystallographic orientation is referred to as a grain.
Each grain typically grows from a single solid seed and presents a distinctive growth direction.
Grain boundaries (GBs) thus emerge during the growth competition between different grains with different pattern migration dynamics.
The grain texture --the resulting macroscopic distribution of grain orientations and GBs-- has a key influence on the material strength~\cite{Hirth72}.

The relative orientation of grains is known to be a key determinant of the grain texture.
The classical theory of GB evolution during polycrystalline solidification assumes that grains that have a crystalline orientation best aligned with the   temperature gradient direction will progressively eliminate less aligned neighboring grains during growth~\cite{WaltonChalmers59, Esaka86Thesis}.
Unexpected phenomena, however, such as overgrowth of favorably-oriented grains by unfavorably-oriented grains have been reported experimentally~\cite{WagnerEtAl04, Zhou08} and more recently studied theoretically and computationally. While those simulations have revealed a much more complex picture of grain competition than initially anticipated, they have been confined to 2D or quasi-2D geometries~\cite{LiWang12, TakakiEtAl14, Tourret17}. 
In spatially extended 3D samples, the geometry of practical relevance, this competition has remained largely unexplored, for two main reasons.
First, gravity-induced buoyancy in the liquid phase~\cite{Mehrabian70,Dupouy89,Jamgotchian01} has made it traditionally difficult to solidify samples under well-controlled homogeneous conditions, a requirement to elucidate grain competition mechanisms inherent to the non-equilibrium growth process.
Second, simulations of alloy solidification in 3D spatially extended samples have remained computationally costly~\cite{Shimokawabe11}.

Herein, we address these challenges by combining experiments~\cite{BergeonEtAl13,Mota15, AkamatsuNguyen16, PeredaEtAl17, MotaEtAl21} performed in reduced gravity conditions onboard the International Space Station (ISS) with massively-parallelized quantitative phase-field (PF) simulations~\cite{TourretEtAl15,TourretKarma15, Tourret17,SongEtAl18:ActaMat, KurzFisherRohit19, KurzRappazTrivedi21}.
Our results shed light on the complex behavior of 3D GBs and their stability during polycrystalline solidification.


\section*{Results and Discussion}

Experiments were performed in the Directional Solidification Insert (DSI) of the DEvice for the study of Critical LIquids and Crystallization (DECLIC) onboard the ISS.
This setup allows imaging  {in situ} the evolution of a solid-liquid interface during directional solidification of transparent organic compounds.
The reduced gravity permits well-controlled and homogeneous conditions, with predominantly diffusive transport of heat and solute species, even in bulk 3D samples~\cite{BergeonEtAl13,TourretEtAl15,Mota15,PeredaEtAl17, MotaEtAl21}.

\begin{figure}
\centering
\includegraphics[width = 180 mm]{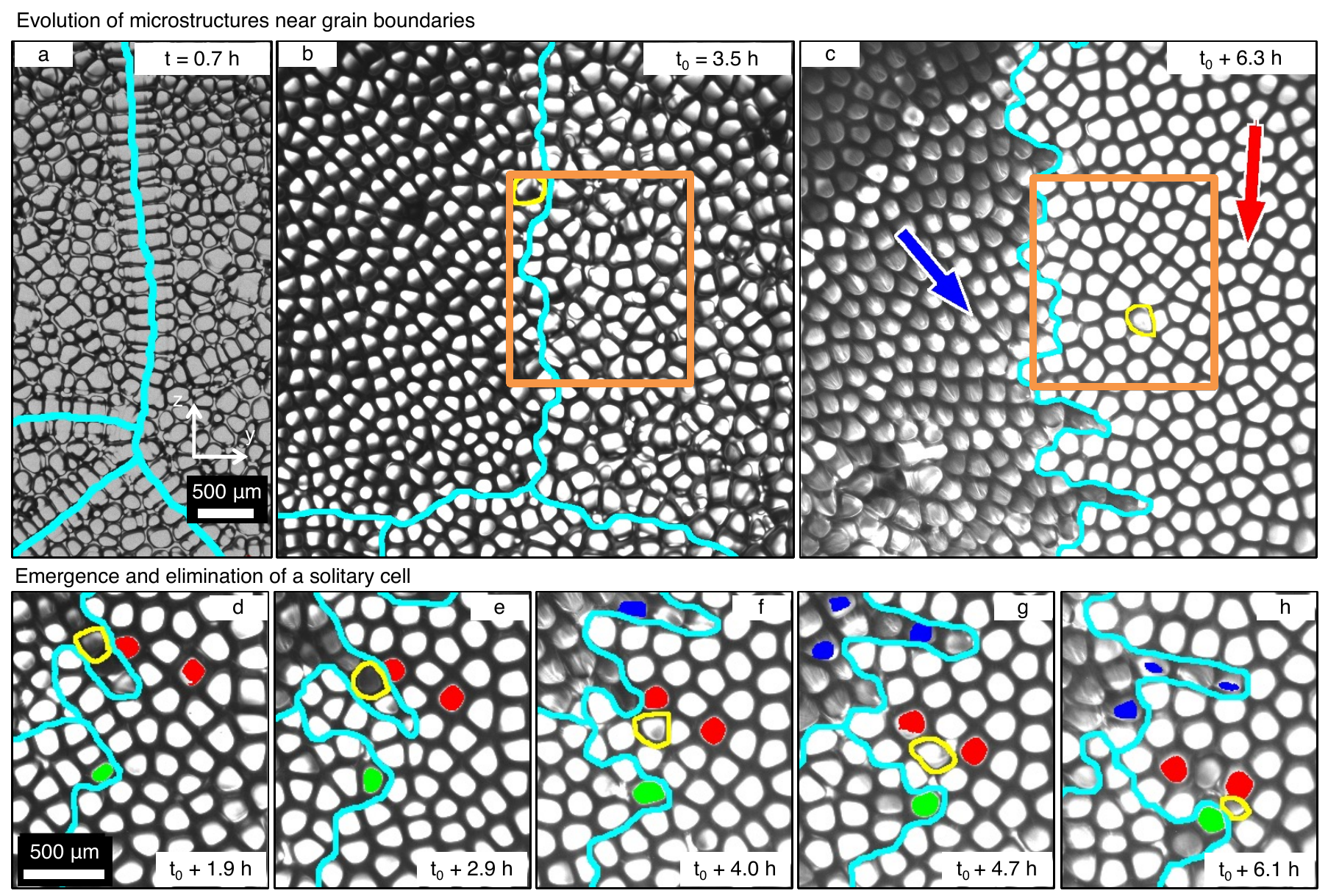} 
\caption{ 
\rm
\label{fig:exp}
{\bf Evolution of the solid-liquid interface pattern, grain boundaries, and emergence of solitary cells.}  
The interface is seen from its liquid side during directional solidification experiments in microgravity at different times $t$. 
Grain boundaries (GBs) appear in cyan lines. 
The GBs have a uniform linear morphology at early time in {\suba}.
A solitary cell (SC), outlined in yellow, emerges in the orange square of {\subb}-{\subc} (also shown in Fig.~\ref{fig:solitaryexp}). 
Red and blue arrows in {\subc} indicate the overall drift directions of the grains. 
Another SC (outlined in yellow) in panels {\subd}-{\sube} emerges at $t_0+4.0$~{\uth} {\subf} and survives in a short time g before being eliminated {\subh}. 
See Supplementary Movie 1 of {\suba}-{\subc}, and 2 of {\subd}-{\subh}. 
}
\end{figure}

Figure~\ref{fig:exp} shows optical images of the solid-liquid interface during the growth of a transparent succinonitrile-0.24wt\%~camphor compound at an imposed growth velocity $V=1.5$~{\utv} within a temperature gradient $G = 19$~{\utG}, seen from a camera placed on the top of the liquid and directly facing the advance of the solidification front. 
Shortly after the start of the experiment, the undercooled planar solid-liquid interface destabilizes and GBs appear, which are most noticeable from the distinct drifting directions of the different grains (Supplementary Movies 1-3).
These GBs, almost straight initially (Fig.~\ref{fig:exp}{\suba}), progressively develop finger-like protrusions when groups of cells penetrate the cellular pattern of the neighbor grain (Fig.~\ref{fig:exp}{\subb}-{\subc}).
This first {in situ} observation of such morphological roughening of the GB provides a visual explanation for grain morphologies previously observed in post-mortem characterization of fully solidified dendritic metallic microstructures~\cite{YangEtAl13}.

\begin{figure}
\centering
\includegraphics[width = 180 mm]{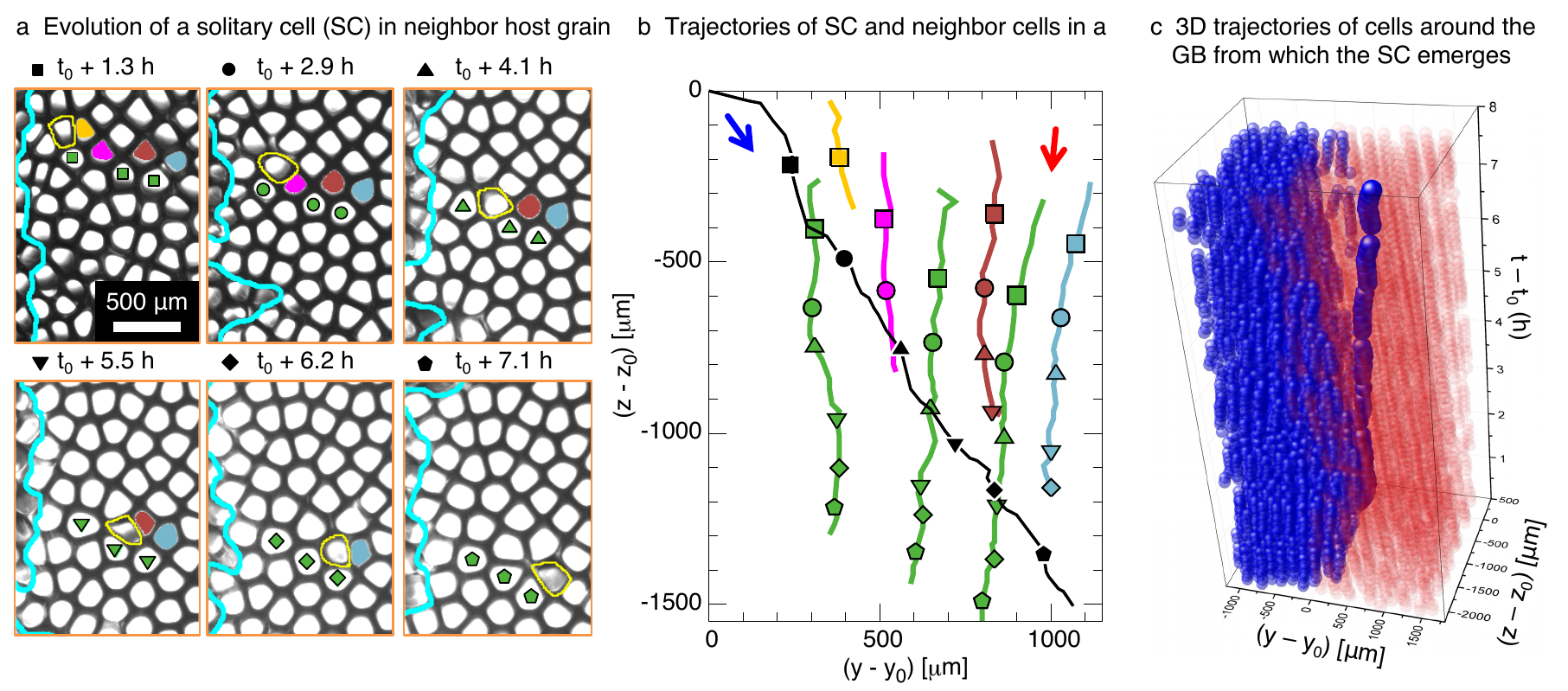} 
\caption{ 
\rm
\label{fig:solitaryexp}
{\bf Emergence and evolution of a solitary cell (SC) in DECLIC-DSI experiments.} 
Snapshots in {\suba} illustrate the evolution of a SC (outlined in yellow) and grain boundary (GB) morphologies (cyan lines).
Trajectories of selected cells (black for the SC; green and other colors for tagged non-eliminated and eliminated host cells, respectively) appear in {\subb} with different symbols as time markers.
Blue and red arrows in {\subb} indicate the overall drift directions of grains. 
Panel {\subc} shows cell trajectories in 3D, evidencing the GB roughening and branching and the resulting tubular defect inside the red grain.
See Supplementary Movies 3 of {\suba}, and 5 of {\subc}. 
}
\end{figure}

A striking observation in these experiments is that the morphological instability of GBs may lead to grain interpenetration and to the invasion of a neighboring grain by a solitary cell (SC) after it detaches from its original grain (outlined in yellow in Figs.~\ref{fig:exp}-\ref{fig:solitaryexp}).
Several events of grain interpenetration and SC invasion were observed at a growth velocity $V = 1.5~$\textmu m/s (Figs.~\ref{fig:exp}-\ref{fig:solitaryexp} and Supplementary Movies 1-3) but also $V = 2~$\textmu m/s (Supplementary Figs.~1-2 and Supplementary Movie 4). 
SCs typically emerge when a penetrating group of cells becomes disconnected from its original grain through the elimination of cells at the base of a finger-like protrusion by cells of the penetrated grain. 
Once detached from their original grain, SCs drift within the nearby (host) grain for several hours of experiments (Fig.~\ref{fig:exp}\suba-\subc), but they can occasionally be eliminated, for instance when squeezed between two cells of different grains (Fig.~\ref{fig:exp}{\subd}-{\subh}).


Figure~\ref{fig:solitaryexp} illustrates the time evolution of a SC within its host cellular pattern, namely within the orange box region of Fig.~\ref{fig:exp}{\subb}-{\subc}.
The SC emerges at a time $t = t_0$ (Fig.~\ref{fig:exp}{\subb}) and drifts within its host grain until the end of the experiment at $t=t_0+7.6$~{\uth} (Fig.~\ref{fig:solitaryexp}{\suba}).
The observed trajectories of the SC and selected cells within the host grain appear in Fig.~\ref{fig:solitaryexp}{\subb}, respectively in black line and colored lines corresponding to tagged cells in {\suba}. 

Cells within the host grain move in an overall consistent direction (red arrow), but the motion of the SC remains prescribed by its own crystalline orientation (blue arrow).
The evolution of the SC involves the elimination of cells within the host pattern (non-green trajectories in Fig.~\ref{fig:solitaryexp}{\subb}) as well as accommodating the deformation of cells as they pass by each other without elimination (green trajectories).
The drifting of a SC leads to the formation of an unexpected 3D branched structure consisting of a cylindrically shaped branch of approximately one cell diameter embedded in the host grain, as illustrated by the 3D reconstruction in Fig.~\ref{fig:solitaryexp}{\subc} (Supplementary Movie 5). 
Such a tubular defect in the grain structure could have a major impact on the mechanical behavior of the microstructure.


We used a quantitative 3D PF model for dilute alloy directional solidification~\cite{Echebarria04,TourretKarma15, Tourret17} to explore the mechanism of 3D branching of GBs (Fig.~\ref{fig:exp}{\subc}) linked to the SC emergence.
We simulated bi-crystalline microstructures that grow under the experimental conditions, i.e. considering a SCN-0.24wt\% camphor alloy growing at $V = 1.5~${\utv} and $G = 19$~{\utG}. Detailed simulation parameters are provided in Supplementary Methods. 
The only difference between the two grains is their crystal orientations. Each grain has a $\langle100\rangle$ preferred growth direction aligned along the axis $x'''$ (see below). 
This $\langle100\rangle$ direction makes an angle $\theta$ with respect to the $x$ axis (parallel to the $G$ direction), which can be inferred from the drift velocity of cells within a grain (see details in Supplementary Methods). The drift direction of those cells, as observed in the experimental movies, is defined by the projection angle $\phi$ of the $x'''$ axis onto the $y$-$z$ observation plane. 
We use $\phi_1$ and $\phi_2$, where the subscript stands for the two different grains (see below). 
Angles $\phi_1$ for grain 1 (blue) and $\phi_2$ for grain 2 (red) are measured from the $y+$ and $y-$ directions, respectively. Those drift angles were identified as key parameters for the occurrence of the invasion events observed in the experiments. 
The detailed relation between ($\theta$, $\phi$) angles discussed here and the equivalent Euler angles ($\alpha$, $\gamma$) used as input parameters to the phase-field simulations are fully detailed in the attached Supplementary Methods.
As discussed below, these simulations allowed identifying the critical importance of the crystalline orientations and of the primary array spacing (see Supplementary Notes 2-3) upon the morphological stability of the resulting GB.

\begin{figure}
\centering
\includegraphics[width = 180 mm]{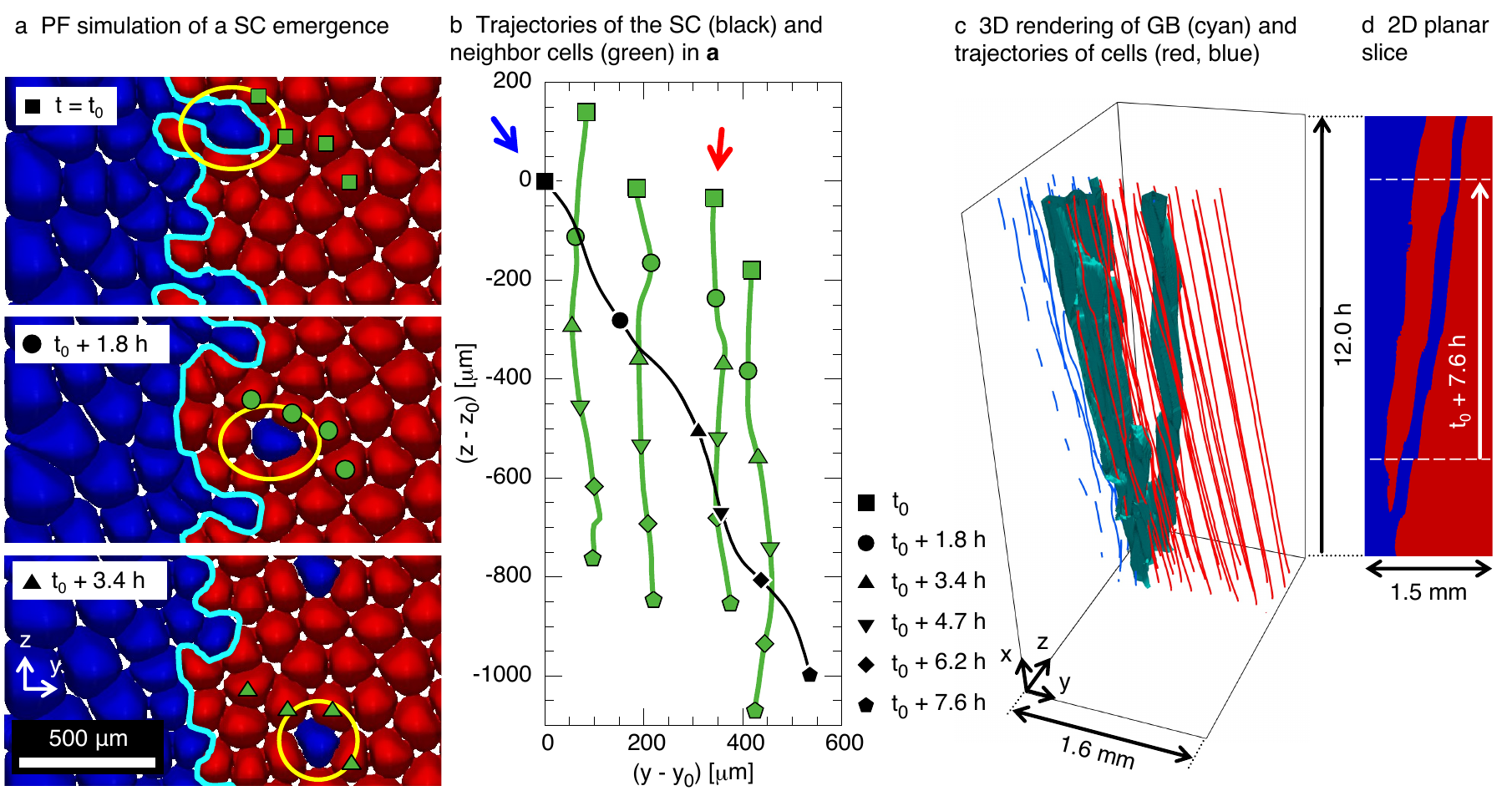} 
\caption{ 
\rm
\label{fig:solitarysim}
{\bf Emergence and evolution of a solitary cell (SC) in 3D PF simulation of DECLIC-DSI experiments.} 
Panel {\suba} (Supplementary Movie 6) shows the simulated pattern evolution for experimental parameters and grain orientations. 
For clarity, part of the images is duplicated vertically using the periodic boundary conditions. 
The emerging SC is outlined in yellow. 
Trajectories of selected cells (green) and the SC (black) over $7.6$~{\uth} appear in {\subb}, with symbols as time markers. 
Color arrows indicate drift directions of the blue and red grains. 
The 3D reconstruction of cell trajectories and the grain boundary (GB) in {\subc} (Supplementary Movie 7) and the 2D planar section in {\subd} show the progressive roughening and branching of the GB.
Grain orientations are $(\theta_1,\phi_1)=(6^\circ, -56^\circ)$ and $(\theta_2,\phi_2)=(3^\circ, 84^\circ)$ as extracted from the analysis of the experimental videos (see details in Supplementary Methods).
}
\end{figure}

First, we performed a spatially-extended bi-crystalline simulation with similar conditions and crystal orientations as in the experiment of Figs.~1-2
, i.e. with $(\theta_1,\phi_1)=(6^\circ, -56^\circ)$ and $(\theta_2,\phi_2)=(3^\circ, 84^\circ)$ (see details on the identification of crystalline orientations from Supplementary Movies 1-3).
Figure~\ref{fig:solitarysim} shows the results of a simulation that started with a straight GB normal to the $y$-direction. 
Like in the experiment, the GB (cyan line in Fig.~\ref{fig:solitarysim}{\suba}) is morphologically unstable, as cells in the left (blue) grain tend to penetrate into the right (red) grain.
The leader cell of an invading group, outlined in yellow in Fig.~\ref{fig:solitarysim}{\suba}, detaches and becomes a SC at $t = t_0$.
Then, it progresses within the host grain for over $7.6$~{\uth} (Supplementary Movie 6), like in the experiment. 
The trajectories of cells in the host grain (green lines and symbols in Fig.~\ref{fig:solitarysim}{\subb}) are consistent with the grain crystalline orientation (red arrow).
Meanwhile, the SC trajectory (black line and symbols) follows its natural drifting direction dictated by the crystal orientation of its parent grain (blue arrow).
The 3D mechanism of GB branching linked to SC emergence is illustrated in Fig.~\ref{fig:solitarysim}{\subc}, with a 2D slice through a plane containing the SC trajectory in Fig.~\ref{fig:solitarysim}{\subd} (Supplementary Movie 7). 
The resulting 3D roughening and branching of the GB are caused by the morphological instability of the GB and the subsequent SC emergence. 

Experiments (Fig.~\ref{fig:solitaryexp}) and PF simulations (Fig.~\ref{fig:solitarysim}) are in quantitative agreement with regards to the drifting direction and velocity of the SC. 
Yet, one interesting difference is that, in the experiments, the SC eliminates cells from the invaded grain along its path, while the SC  in the PF simulation does not produce any cell elimination. 
We attribute this difference to the lower average cell spacing in experiments compared to PF simulations, in part due to the simplified frozen temperature profile assumption~\cite{SongEtAl18:ActaMat}.
The lower cell spacing in experiments renders cells inside the invaded grain more susceptible to elimination as the local spacing fluctuates to accommodate the passage of the SC. 
This interpretation is supported by additional PF simulations that relate the SC fate to the local array spacing (see Supplementary Figs.~8-9 and Supplementary Note 3).

\begin{figure}
\centering
\includegraphics[width = 88 mm]{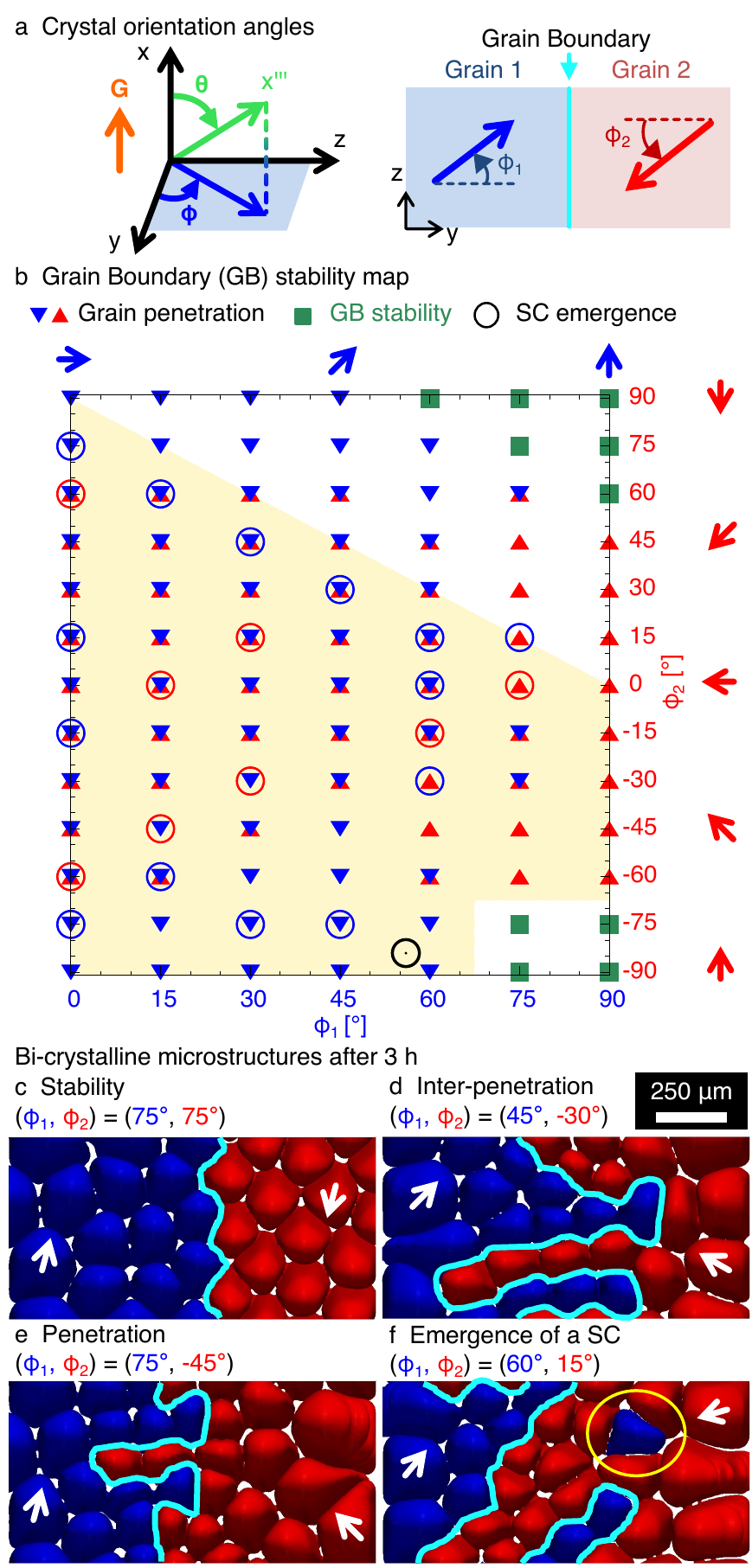} 
\vspace{-30pt}
\caption{ 
\rm
\label{fig:GC3D}
{\bf Link between the stability of grain boundaries (GBs) and the crystal orientations.} 
Panel {\suba} shows the definition of crystal orientation angles $\theta$ and $\phi$. 
Subscripts 1 and 2 correspond to the left (blue) and the right (red) grains, respectively. 
Panel {\subb} (and Supplementary Fig.~7) shows the mapping of the convergent GB stability with $\theta_1=\theta_2=5^\circ$, which results in GB stability (green squares), grain penetration (triangles of the color of the penetrating grain), and/or solitary cell (SC) emergence (circles of the color of the emerging SC). 
The experimental SC is shown as the black circle. 
Arrows in {\subb}-{\subc} indicate drifting directions of grains.
Panels {\subc}-{\subf} show representative microstructures after $3$~{\uth} (Supplementary Movie 8). 
}
\end{figure}

Experiments only produce few large grains with different misorientations, and hence a small number of different GB types.
Thus, we used PF simulations to explore more systematically the role of GB bi-crystallography on the morphological stability of GBs. 
The results shown in Fig.~\ref{fig:GC3D} demonstrate that grain interpenetration and SC invasion occur over a wide range of misorientation angles that characterize the GB bi-crystallography. 

Simulations were started with a straight GB normal to the $y$-direction, keeping a similar crystal tilt angle with respect to the temperature gradient $\theta = 5^\circ$ for both grains (see definition of angles in Fig.~\ref{fig:GC3D}\suba) and varying the azimuthal angles $\phi_1$ and $\phi_2$, corresponding to the pattern drift directions. 
We selected a low $\theta$ value because: (i) only low tilt angles are observed experimentally, e.g. $\theta \approx 6^\circ$ and $3^\circ$ for the left and right grains in Fig.~\ref{fig:exp}, respectively (see the method section and Supplementary Methods), and (ii) we found that GBs are usually morphologically stable for a large $\theta$. 
In particular, for a convergent GB between a well-oriented grain with $\theta_1=\phi_1=0^\circ$ and a misoriented grain of varying misorientation $\theta_2$ with $\phi_2 = 0^\circ$, simulations reveal that the GB is only morphologically unstable for $\theta_2$ lower than about $15^\circ$ (see Supplementary Fig.~6 and Supplementary Note 2). 
Moreover, while we primarily focused on the influence of the crystal orientations ($\phi_1, \phi_2$) on the GB morphology, we also identified a key influence of the array spacing of initial microstructure upon the GB roughening dynamics. Indeed, cell invasion is promoted by a higher primary spacing within the host grain, as illustrated and discussed within the Supplementary Note 3 (see Supplementary Figs.~8 and 9).

Results of the mapping are summarized in Fig.~\ref{fig:GC3D}{\subb}, with representative microstructures after $3$~{\uth} of growth illustrated in Fig.~\ref{fig:GC3D}{\subc}-{\subf} (also see Supplementary Fig.~7).
The entire competing dynamics of Fig.~\ref{fig:GC3D}{\subc}-{\subf} can be found in Supplementary Movie 8.
Inter-penetration (Fig.~\ref{fig:GC3D}{\subd}) or one-sided penetration (Fig.~\ref{fig:GC3D}{\sube}) of grains, marked as triangles in Fig.~\ref{fig:GC3D}{\subb}, occur over a wide range of orientations. 
However, the invasion by a grain with $|\phi| \ge 75^\circ$ was not observed. 
Hence, when both $|\phi_1|$ and $|\phi_2|$ are high, which corresponds to parallel drifting directions where the two grains slide against each other, the GB remains stable (Fig.~\ref{fig:GC3D}{\subc} and green squares in Fig.~\ref{fig:GC3D}{\subb}).
When grain penetration occurs, within the host grain, the leader cell of a penetrating group can either eliminate cells on its path or accommodate to squeeze itself between cells of the host array. After the leader cell clears a pathway, other penetrating cells follow, which results in the morphological destabilization of the GB. Following destabilization and penetration, this pathway may be interrupted by the elimination of one of the trailing cells by those of the host grain, hence leading to the isolation of an individual cell or group of cells. 
Events of SC emergence (Fig.~\ref{fig:GC3D}{\subf} and circles in Fig.~\ref{fig:GC3D}{\subb}) are only observed for $\phi_1 + \phi_2 \le 90^\circ$ when at least one grain penetrates into the other grain, i.e. for both $|\phi_1|$ and $|\phi_2|$ lower than $75^\circ$.
This region (yellow background in Fig.~\ref{fig:GC3D}{\subb}) contains the experimental conditions of Fig.~\ref{fig:exp} (black circle).



Even though this fingering instability of GBs was visualized here {in situ} for the first time in a transparent organic alloy, we expect it to occur during solidification of a wide range of technological metallic alloys, thereby potentially impacting mechanical behavior and other properties through the creation of complex 3D GB structures. 
Post-mortem metallographic analysis of Ni-based superalloys has revealed interpenetrating grain structures in serial sections~\cite{YangEtAl13}, thus providing an indirect evidence of this instability in a well-developed dendritic growth regime. 
However, solidifying grains imaged in 2D through serial sectioning are usually reported to remain connected regions of space, i.e. with a GB evolution typically occurring by the progressive invasion of a pattern into its neighbor. 
Rapid advances in 3D and 4D (time-resolved 3D) imaging of metallic systems, e.g. using X-ray tomography, should yield further insight into the mechanisms of roughening, fingering, and solitary cell or dendrite emergence in technological metallic alloys.
Beyond the field of solidification, we expect the present results to have potential relevance for the interpenetration of drifting cellular patterns in other contexts including biological systems.


\section*{Methods}

\subsection{Microgravity experiments}

Experiments were performed within the Directional Solidification Insert (DSI) of the Device for the study of Critical LIquids and Crystallization (DECLIC) aboard the International Space Station.
Reduced gravity allows the processing of homogeneous 3D cylindrical samples using a classical Bridgman setup with various pulling velocities $V$ and temperature gradients $G$. 
The use of a transparent succinonitrile (SCN)-0.24wt\%~camphor alloy permits {in situ} optical imaging of the interface~\cite{Mota15}.
Experiments are typically initiated from a single crystal growing along the temperature gradient direction. 
Yet, after repeated experimental cycles of solidification and remelting of a given sample, polygonization of the seed may progressively develop, ultimately leading to the appearance of grain boundaries (GBs)~\cite{PeredaEtAl17, MotaEtAl21}. 

The alloy solidifies within a cylindrical crucible with an inner diameter of $10$~{\utmm}. 
The end temperatures of the setup are set higher and lower than the liquidus temperature of the alloy, such that the solid-liquid interface is located within a stable and well-controlled temperature gradient. 
The crucible is pulled through an axial adiabatic zone, allowing the interface to grow up to $100$~{\utmm} in length at a controlled velocity.
Thus, we can investigate the microstructure selection and image the {in situ} evolution of the solid-liquid interface while independently controlling the growth velocity $V$ and temperature gradient $G$ experienced by the interface.

Here we focus on the experiment with $G = 19$~{\utG} and $V = 1.5$~{\utv} (and $V = 2.0$~{\utv} in Supplementary Figs.~1-2), which yields a morphological destabilization of the grain boundary and the emergence of solitary cells.  
Further details on the experimental setup appear in previous publications~\cite{BergeonEtAl13, TourretEtAl15, Mota15, PeredaEtAl17, MotaEtAl21}.
\newline 

\subsection{Phase-field simulations}

We used a quantitative phase-field formulation for directional solidification~\cite{Echebarria04, Echebarria10}. 
The model assumes a one-dimensional frozen temperature profile, with a set temperature gradient $G$ moving at a constant velocity $V$ with respect to the sample, neglecting solute diffusion in the solid phase. 
The model was developed for exact asymptotic matching of the corresponding solid-liquid sharp-interface problem, even using a diffuse interface width much larger than its actual physical size~\cite{Echebarria04, KarmaRappel98}.
The time evolution equation for the solute field includes an anti-trapping current across the solid-liquid interface in order to correct for the spurious numerical solute trapping arising from the use of a such a wide diffuse interface~\cite{Karma01,Echebarria04}.
The model only uses one phase field for the two grains.
The orientation of each grain is set by introducing an integer grain index field which prescribes the three-dimensional orientation of the grain~\cite{TourretKarma15}.

Current simulations focus on one solidification condition, namely $V = 1.5~${\utv} and $G = 19$~{\utG}, for a SCN-0.24wt\%~camphor alloy.
Material and numerical parameters are similar to those used and discussed in previous publications~\cite{SongEtAl18:ActaMat,PeredaEtAl17} (Supplementary Table 2). 
The problem consists of two coupled partial differential equations, which are solved using finite differences on a homogeneous grid of cubic elements. 
We use a non-linear preconditioning of the phase field~\cite{Glasner01}, hence solving for the preconditioned phase field $\psi$ defined as $\varphi = \tanh ( \psi / \sqrt{2})$, where $\varphi$ is the standard phase field that varies smoothly from $\varphi = +1$ in the solid to $\varphi = -1$ in the liquid. This simple nonlinear change of variable enhances the numerical stability of the simulations for larger grid spacings, and therefore allows faster (or larger) simulations with negligible loss in accuracy~\cite{Glasner01,BergeonEtAl13, TourretEtAl15}. Detailed equations and relations between $\psi$, $\varphi$, and the two coupled partial differential equations are provided in Supplementary Methods as well as in previous works~\cite{Glasner01,BergeonEtAl13, TourretEtAl15, SongEtAl18:ActaMat}.
For numerical parameters, we use a diffuse interface thickness $W = 2.7$~{\utmicron}, a homogeneous grid of cubic elements of size $\Delta x = 3.2$~{\utmicron}, and an explicit time step of $\Delta t = 5.7~{\rm ms}$. 
Thermal fluctuations are mimicked by the introduction of a random noise perturbation onto the preconditioned phase field $\psi$, with a dimensionless amplitude $F_{\psi}$ = 0.01~\cite{TourretEtAl15}. 
Simulations were parallelized on Graphics Processing Units (GPUs) using Nvidia\textsuperscript{TM} CUDA programming platform.

The simulation domain size was $L_x \times L_y \times L_z~$[\textmu m$^3] = 2295 \times 1912 \times 633$ for Fig.~\ref{fig:solitarysim}{\suba} and $2295 \times 1272 \times 633$ for Fig.~\ref{fig:GC3D}{\subb}, where $L_x$, $L_y$, and $L_z$ respectively correspond to spatial directions $x$, $y$, and $z$. 
We use periodic conditions along the $z$-normal boundaries, and no-flux symmetric conditions for all other boundaries. 
The GB between the two grains is initially planar and normal to the $y$-direction. 
The GB of the simulation in Fig.~\ref{fig:solitarysim}{\suba} is initially located at a distance $L_y/4$ from the left ($y=0$) boundary.
This simulation over $t_{max} = 12$~{\uth} was performed in about 12 days ($\approx 290$ hours) using eight Nvidia Tesla K80 GPUs. 
In Fig.~\ref{fig:GC3D}{\subb}, the GB is initially located at the center of the domain in the $y$ direction. 
These simulations were performed for a total solidification time $t_{max} = 3$~{\uth}. 
Each one of the 69 simulations took between 150 and 170 hours to perform on one Nvidia GeForce GTX Titan X GPU.

\subsection{Crystal orientation and drift velocity}

When a crystal is misoriented with respect to the temperature gradient by an angle $\theta$, the resulting cellular or dendritic array drifts laterally with a lateral velocity $V_d$, projected on the experimental observation plane.
Previous studies using a thin-sample geometry~\cite{Akamatsu97, Deschamps08} have shown that the drift angle $\theta_d=\tan^{-1}({V_d}/{V})$ is typically lower than the crystal orientation $\theta$.
The ratio between $\theta_d$ and $\theta$ is a function of the P\'eclet number $\Pe = \lambda V/D$, where $D$ is the liquid solute diffusivity (here $D = 270~$\textmu m$^2$/s) and $\lambda$ is the array primary spacing, as:
\begin{equation}
\label{eqn:driftangles}
\frac{\theta_d }{ \theta } = 1 - \frac{1}{1 + f \, \Pe^g } .
\end{equation}
Parameters $f$ and $g$ depend upon the material.
It was initially suggested\cite{Akamatsu97, Deschamps08} that $g$ be a constant while $f$ vary as a function of the grain angle $\theta$.
However, we found good agreement with PF predictions throughout a wide range of orientations keeping both $f$ and $g$ constants~\cite{TourretKarma15}.

We ascertained the validity of Eq.~\eqref{eqn:driftangles} in the current simulations and identified parameters $f = 0.67$ and $g = 1.47$ (see Supplementary Fig.~5 and Supplementary Note 1).
To do so, we simulated hexagonal patterns (Supplementary Fig.~4a) with different primary spacings from $\lambda = 141$ to $320$~{\utmicron}, with a screening step of about 22~{\utmicron}.
On these structures, we imposed crystal orientations of $\theta = 5$, $10$, and $15^\circ$ at $\phi = 0^\circ$ (see Supplementary Fig.~5a).
We also considered $\phi = 30^\circ$ at $\theta = 10^\circ$. 
To calculate $f$ and $g$ from these simulations, we measured the time evolution of the positions of the centers of cells in the $y$-$z$ plane every 333~s using a similar image processing method as described in a previous article~\cite{TourretEtAl15}. 
We measured the drift velocity $V_d$ and thus the angle $\theta_d$, and then fitted simulation results to Eq.~\eqref{eqn:driftangles} for $f$ and $g$ (Supplementary Fig.~5d).
We verified that this relation linking grain orientation and drift velocity was not only valid for hexagonal patterns (Supplementary Figs.~4a, 5a, and 5d), but also held for fcc-like arrays and pseudo-2D thin-sample confined arrays (Supplementary Figs.~4b-c, 5b-c, and 5e), as well as for different azimuthal orientation angles $\phi$ (Supplementary Fig.~5f), using the same values for $f $ and $g$.

Using Eq.~\eqref{eqn:driftangles} with these identified values of $f$ and $g$, we calculated the 3D crystal orientation of different grains in the experiments from the observed drifting velocities and directions.
At $V = 1.5$~{\utv}, the average primary spacing of the overall cellular array is $\lambda \approx 260$~{\utmicron}.
The drift velocities are $V_d = 0.09$ and $0.04$~{\utv} for the left-side and right-side grains in Fig.~\ref{fig:exp}{\subc}, which yields $\theta_d = 3.4^\circ$ and $1.5^\circ$, respectively.
Thus using Eq.~\eqref{eqn:driftangles} with the estimated constants, we calculate crystal angles $\theta = 6^\circ$ for the left-side grain and $3^\circ$ for the right-side grain. 
We then used these values for the imposed crystal angles in the PF simulation of Fig.~\ref{fig:solitarysim}{\suba}.

\section*{Data Availability}
The authors declare that the data supporting the findings of this study are available within the paper and its supplementary information files. 

\section*{Code Availability}
The code for phase-field simulations is available from the corresponding author upon request. 

\section*{References}



\newpage
\begin{addendum}
 \item 
This study was supported by the National Aeronautics and Space Administration Grants NNX16AB54G, 80NSSC19K0135 and 80NSSC22K0664 (Y.S., K.J., R.T., A.K.), and the French Space Agency CNES project MISOL3D Microstructures de solidification 3D (F.M., B.B., N.B.).
D.T. acknowledges support from the Spanish Ministry of Science via a Ramon y Cajal fellowship (RYC2019-028233-I).
Y.S. acknowledges that this work was partially prepared by LLNL under Contract DE-AC52-07NA27344.
Authors also gratefully acknowledge the access to the supercomputing Discovery cluster provided by Northeastern University. 

 \item[Author Contributions Statement] 
Y. S. and A. K. designed the simulations. 
Y. S., D. T., and K. J. implemented the simulation code.
Y. S. and K. J. performed PF simulations. 
Y. S., D. T. and A. K. analyzed simulation data.
F. M., N. B., B. B., and R. T. designed and remotely performed the DECLIC-DSI experiments. 
F. M. and N. B. analyzed experimental data. 
Y. S., D. T., F. M., K. J., N. B., and A. K. wrote the manuscript. 
All authors discussed the results and edited the manuscript. 

 \item[Competing Interests Statement] 
The authors declare that they have no competing financial interests.


\end{addendum}


\pagebreak

\section*{Figure Legends/Captions}

\pagebreak

\pagebreak

\pagebreak


\end{document}


\title{{Cell invasion during competitive growth of polycrystalline solidification patterns}
 \vspace{-5px}
 \\
 \noindent\rule{8 cm}{2.5 pt}
 \\
 Supplementary Information}

\maketitle

\section*{Supplementary Methods}

\subsection{Experimental setup.}

In the Bridgman experimental setup, a linear temperature profile, with a homogeneous temperature gradient of an amplitude $G$ in the direction $x$, is established by controlling a low temperature furnace and a higher temperature furnace on both sides of a thermally insulated (adiabatic) zone. 
At rest, a planar solid-liquid interface forms close to the alloy liquidus temperature, i.e. 330.85~K for a succinonitrile (SCN)-0.24wt\%~camphor alloy. 
When the sample crucible is pulled towards the colder area, the interface initially moves towards the colder area because its growth is slower than the pulling velocity $V$, which is also the target steady state growth velocity. 
During this initial interface recoil, the concentration at the interface on the liquid side increases due to the rejection of camphor by the growing solid of lower camphor solubility than the liquid phase.
The resulting increase of the solute gradient at the interface leads to an increase of the interface velocity. 
Once the interface velocity exceeds a critical value that depends on $G$ and on the alloy phase diagram, the planar interface becomes morphologically unstable to perturbations~\cite{MullinsSekerka64,TillerEtAl53,SongEtAl18:ActaMat}. 
Then, small ridges and poxes appear along the interface (see e.g. Fig.~2 of Ref.~\cite{SongEtAl18:ActaMat} and early stage of Supplementary Movie 1), which ultimately develop into grain boundaries (GBs) and cells, respectively~\cite{Bergeon11, Bergeon15,SongEtAl18:ActaMat}. 
The pattern coarsens as cells compete with one another through thermal and solute fields, until enough cells have been eliminated and a stable array spacing is established among surviving cells. 
It is worth noting that GBs here are all low-angle GBs, which may also be referred to as subboundaries~\cite{Bottin-Rousseau02, MotaEtAl21, Faivre13}.

\subsection{Experimental parameters.}

Various $(G,V)$ conditions were investigated within the DECLIC-DSI. 
In previous studies~\cite{BergeonEtAl13,TourretEtAl15, PeredaEtAl17}, we focused on a narrow range of parameters that resulted in oscillating dynamics of cellular arrays, for instance for $V = 0.5$ to $1.5$~{\utv} at $G=19$~{\utG}.
In the current study, we focus on experiments in which we observed at least one converging GB, that is for $V = 1.5$ and $2.0$~{\utv} at $G$ = 19~{\utG}. 
Material parameters for a SCN-0.24wt\%~camphor alloy are listed in Supplementary Table~\ref{tab:params} and have been discussed elsewhere~\cite{TourretEtAl15, Mota15, Mota16}.

\subsection{Experimental observations.}

\begin{figure}
\centering
\includegraphics[width = 89 mm]{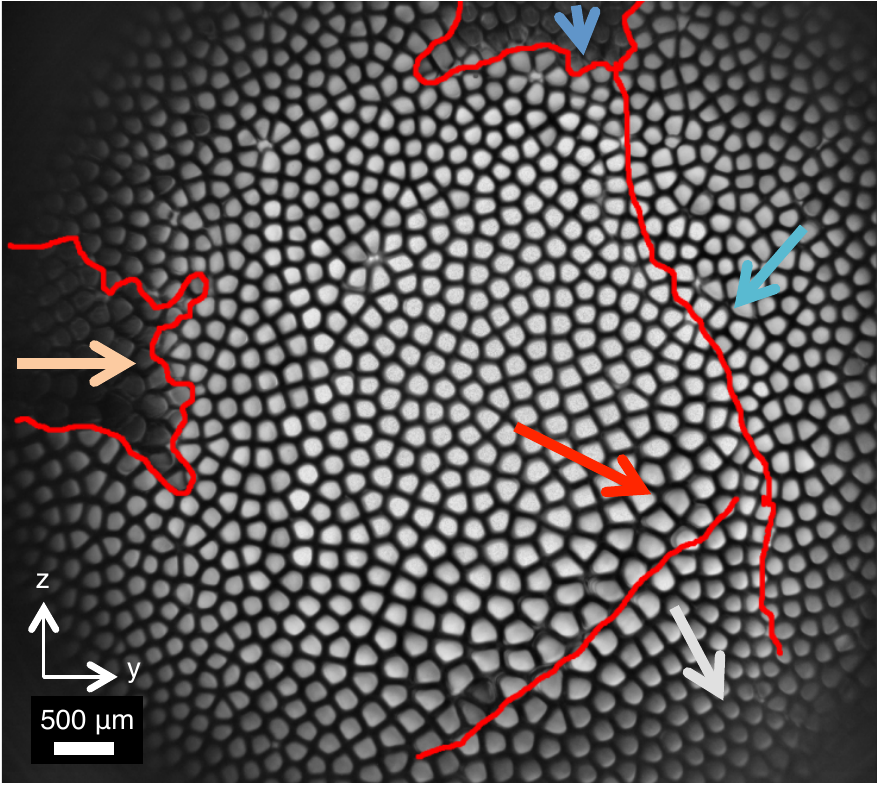}
\\
\caption{ 
\label{fig:steadyexp}
\rm
{\bf Solid-liquid interface pattern of a SCN-0.24wt\%~camphor alloy at $V = 2.0$~{\utv} and $G = 19$~{\utG}.} 
Solid-liquid interface seen from the liquid, facing the main growth direction. A group of cells (hereafter called grain) drifts in a same direction (arrows) due to their crystal orientation, and lateral dynamics of these groups lead to the evolution of boundaries (red lines) between them. 
}
\end{figure}
\begin{figure}
\centering
\includegraphics[width = 183 mm]{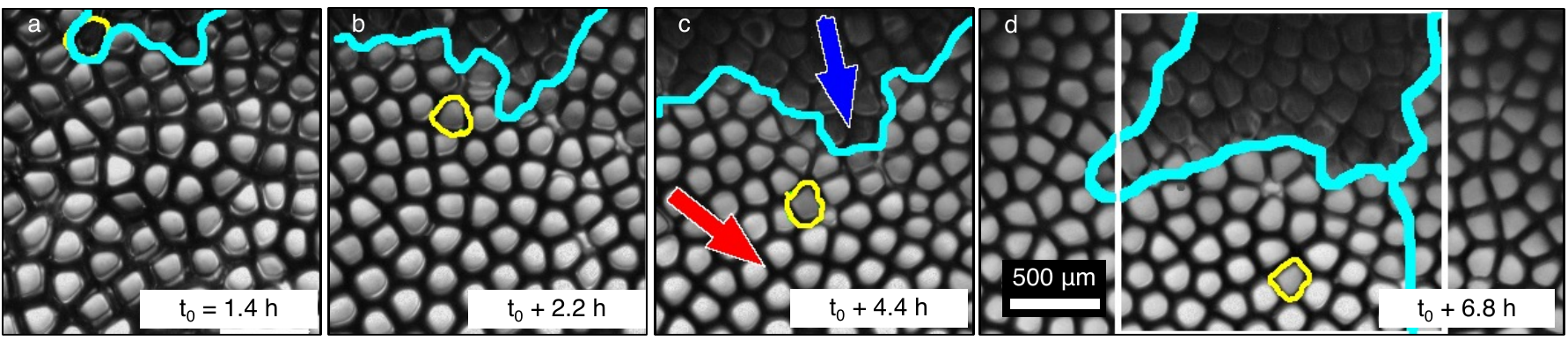}
\caption{ 
\label{fig:solexpv2}
\rm
{\bf Dynamics of a solitary cell (SC) at $V = 2.0$~{\utv} and $G = 19$~{\utG}.}
As also observed in Fig.~2{\suba} of the Letter for $V = 1.5$~{\utv}, a SC emerges in the experiment at $V = 2.0$~{\utv}. 
While the upper grain invades the lower grain (left to right), a cell (circled in yellow) near the convergent GB (cyan line) detaches from its original grain and become a SC ({\suba}). 
This SC survives and drifts within the lower host grain ({\subb}-{\subd}) until the end of the experiment.
Arrows on top of {\subc} indicate the overall drift direction of these grains. 
The white box in {\subd} represents the region of {\suba}-{\subc} at earlier times. 
}
\end{figure}

As shown in Supplementary Movie 1 for $V = 1.5$~{\utv} and Supplementary Fig.~\ref{fig:steadyexp} below for $V = 2.0$~{\utv}, the growing microstructure consists of several cellular grains with different crystal orientations, which can easily be distinguished due to their distinct drifting directions, i.e. their overall growth direction projected in the $y$-$z$ projection plane (illustrated with arrows in Supplementary Fig.~\ref{fig:steadyexp}). 
Thus, GBs (red lines in Supplementary Fig.~\ref{fig:steadyexp}) form between grains.  
When the two different grains grow towards each other, the GB is called converging, and cells directly adjacent to the GB progressively eliminate one another by impingement~\cite{LuHunt92, HuntLu96, TourretKarma15,TakakiEtAl16:ISIJ, LiWang12, Tourret17}. 
These successive elimination events give rise to the GB roughening and its complex morphologies (e.g. Supplementary Fig.~\ref{fig:steadyexp}). 
Supplementary Movies 2 and 3 show a SC emerging at $V = 1.5$~{\utv}. 
We also observe emergence of a SC for $V = 2.0$~{\utv} and $G=19$~{\utG} (Supplementary Movie 4 and Supplementary Fig.~\ref{fig:solexpv2}).
Whenever a SC emerges, it drifts within its host grain up to several hours following its own crystal orientation (Supplementary Fig.~\ref{fig:solexpv2}{\subb}-{\subd}). 
This continuous SC growth leads to GB branching and tubular defects (Figs.~2c and 3c in the Letter). 


\subsection{Post-processing of experimental results.}

Shortly after the pulling of the sample starts, the planar solid-liquid interface destabilizes (see early stage of Supplementary Movie 1), and the primary spacing $\lambda$ in cellular arrays increases due to the growth competition among cells~\cite{SongEtAl18:ActaMat,PeredaEtAl17}. 
An initial transient oscillatory growth regime follows~\cite{PeredaEtAl17,BergeonEtAl13,TourretEtAl15}.
We discard this transient regime in our estimation of crystalline orientations by performing our analysis exclusively on well-developed microstructures, growing as an overall solidification front at a steady undercooling and velocity, i.e. after $t \approx 3.5$~{\uth} for $V = 1.5$~{\utv} or $t \approx 1.4$~{\uth} for $V = 2.0$~{\utv}.
We use image processing software {\it Visilog} and {\it ImageJ} to extract the center locations in the $y$-$z$ plane for every cell, as well as their evolution in time~\cite{TourretEtAl15,PeredaEtAl17}.
We then extract the drifting velocity for each cell, $V_d$, which allows us to: (1) group them into different grains, and (2) estimate the three-dimensional crystalline orientation of each grain.
Given that the observed microstructures are growing at a steady velocity $V$ along the $x$-direction, as imposed by the pulling velocity of the crucible, we can extract the main growth direction angle $\theta_d$ with respect to the temperature gradient (and pulling) direction $x$ with the simple geometrical relation 
{\begin{equation}
\label{eqn:driftangle}
\tan \theta_d = V_d/ V .
\end{equation}
}%
The growth direction $\theta_d$ usually differs from the crystal orientation $\theta$~\cite{Akamatsu97,Deschamps08,LiWang12,GhmadhEtAl14,TourretKarma15}. 
The relation between a growth angle $\theta_d$ and a crystal orientation angle $\theta$ is discussed later in this document, in light of established theory and additional thorough phase-field validation. 
In Fig.~2 of the Letter with $V = 1.5$~{\utv}, we measured $V_d = 0.09$~{\utv} and $\theta_d = 3.4^\circ$ for the left grain, 
and $V_d = 0.04~${\utv} and $\theta_d = 1.5^\circ$ for the right grain. 
In Fig.~1{\subd}-{\subh} in the Letter, the lower-left grain had $V_d = 0.07$~{\utv} and $\theta_d = 2.5^\circ$, and other grains were similar as in Fig.~2 of the Letter.


\subsection{Simulation model.}

Here we use a PF model that derives from a thin-interface asymptotic analysis, such that it reduces to the corresponding sharp-interface problem while representing the solid-liquid interface as a diffuse interface with a width that is much larger than the actual physical thickness of the interface~\cite{KarmaRappel96,KarmaRappel98}. 
The model neglects solid-state diffusion of solute because it is much slower in the solid than in the liquid phase.
Hence, accounting for a predominantly diffusive transport regime in microgravity conditions, solute transport occurs through pure Fickian diffusion in the liquid phase.
The model includes an anti-trapping term that corrects the spurious solute trapping due to the width of the diffuse interface~\cite{Karma01,Echebarria04}.
The model is quantitative within a dilute binary alloy limit, such that the phase diagram of the solid-liquid interface equilibrium can be described using constants for the interface solute partition coefficient $k$ and the alloy liquidus slope $m$.
The resulting model, briefly described below, is essentially similar to that described in Refs.~\cite{Echebarria04, Echebarria10}, only with a few minor additions to improve its computational efficiency~\cite{TourretEtAl15,TourretKarma15,Tourret17}. 

Since the transport of heat is much faster than the transport of solute, we consider the temperature field as an externally imposed boundary condition.
We assume that the temperature profile is frozen in the setup with a fixed temperature gradient $G$ along the $x$-axis, which results in the common assumption referred to as frozen temperature approximation~\cite{BergeonEtAl13,TourretEtAl15,PeredaEtAl17,TourretKarma15,Tourret17,Echebarria10,Gurevich10,GhmadhEtAl14}. 
The validity and limitations of this approximation, specifically in the context of the DECLIC-DSI experiments, are discussed in Ref.~\cite{SongEtAl18:ActaMat}.
In the frame of the material, moving at a velocity $V$ with respect to the furnace, the temperature $T$ at a position along the $x$ axis is thus given by 
\begin{equation}
T = T_0 + G (x - V t) , 
\end{equation}
where $T_0$ at $x = 0$ is a reference temperature, here chosen to be the solidus temperature of the alloy. 
Within this temperature field, the evolution of the preconditioned phase field $\psi$ and of the normalized solute concentration $U$ with time $t$ are described by~\cite{Echebarria04, Echebarria10, TourretKarma15}
\begin{align}
\label{eqn:psi}
\left[ 1-(1-k)\frac{ \tilde{x} - \widetilde{V} \tilde{t} }{\tilde{l}_T} \right] a_s(\mathbf{n})^2 \frac{\partial\psi}{\partial t} = \vec\nabla \left[a_s(\mathbf{n})^2\right] \vec\nabla\psi
+ a_s(\mathbf{n})^2 \left[  \nabla^2\psi -\varphi\sqrt{2}|\vec\nabla\psi|^2  \right]  \nonumber\\
+ \sum_{q=x,y,z}\left[ \partial_q\left( |\vec\nabla\psi|^2 a_s(\mathbf{n}) \frac{\partial a_s(\mathbf{n})}{\partial(\partial_q\psi)} \right) \right] 
+ \sqrt{2} \left[ \varphi  - \lambda_{c}  (1-\varphi^2) \left( U+ \frac{ \tilde{x} - \widetilde{V} \tilde{t} }{\tilde{l}_T}\right) \right]  ,
\end{align}
\begin{align}
\label{eqn:U}
\Big( 1+k-(1-k)\varphi \Big) \frac{\partial U}{\partial t} =  
\widetilde{D} \; \vec\nabla\cdot\left[ (1-\varphi)\vec\nabla U\right]  
+~\vec\nabla\cdot\left[ \Big(1+(1-k)U\Big) \frac{(1-\varphi^2)}{2} \frac{\partial\psi}{\partial t} \frac{\vec\nabla\psi}{|\vec\nabla\psi|} \right] 
\nonumber\\
+~\Big[ 1+(1-k)U \Big] \frac{(1-\varphi^2)}{\sqrt{2}} \frac{\partial\psi}{\partial t}  ,
\end{align}
where
\begin{align}
U \equiv \frac{1}{1-k} \left[ \frac{ 2 \, c/c_l^0}{ ( 1+k ) - (1 - k) \varphi } - 1 \right] ,
\end{align}
$c_l^0 = c_{\infty}/k$ is the equilibrium concentration on the liquid side of the interface at the reference temperature $T_0$ for a nominal alloy composition $c_\infty$ with $0 < k < 1$, and $\varphi$ is the standard phase-field variable with $\varphi=-1$ ($+1$) for the liquid (solid) before preconditioning, i.e. $\varphi=\tanh(\psi/\sqrt{2})$. 
Space and time are scaled with respect to the diffuse interface width $W$ and the relaxation time $\tau_0$ at $T_0$, respectively~\cite{Echebarria04}. 
Thus, the dimensionless diffusivity $\widetilde{D}$ is given by
\begin{equation}
\widetilde{D} = \frac{D\tau_0}{W^2}=a_1a_2\frac{W}{d_0} 
\end{equation}
with $a_1=5\sqrt{2}/8$ and $a_2=47/75$ from thin-interface asymptotics~\cite{KarmaRappel96, KarmaRappel98}. 
The interface solute capillary length $d_0$ and the coupling factor $\lambda_c$ are 
\begin{equation}
d_0= \frac{ \Gamma }{ |m|c_\infty(1/k-1)  } ,
\end{equation}
and
\begin{equation}
\lambda_{c} = a_1 \frac{W}{d_0} ,
\end{equation}
where $\Gamma$ is the Gibbs-Thomson coefficient of the solid-liquid interface.
The dimensionless thermal length is
\begin{equation}
\tilde{l}_T =\frac{l_T}{W}= \frac{l_T}{d_0}~\frac{1}{W/d_0} , 
\end{equation}
with $\tilde{x}\equiv x/W$ and $\widetilde{V}\equiv V\tau_0/W$.
The term $a_s(\mathbf{n})$ in Eq.~\eqref{eqn:psi} represents the anisotropy of the surface energy, which is discussed in details later in this document (subsection of Crystal orientation in the numerical model).
Finally, in order to incorporate the effect of thermal fluctuations, we add noise by introducing a random perturbation onto the preconditioned phase field evolution equation~\cite{BergeonEtAl13,TourretEtAl15,Tourret17,GhmadhEtAl14}, using a uniform random distribution with a given amplitude $F_{\psi}$.

\subsection{Numerical implementation.}

Eqs.~\eqref{eqn:psi} and~\eqref{eqn:U} are discretized spatially with centered finite differences with a homogenous grid spacing $\Delta x$, and using an explicit Euler time stepping scheme with a constant time step $\Delta t$~\cite{Echebarria04,TourretKarma15,Tourret17,Echebarria10,Gurevich10}. 
For stability of the explicit scheme, we use a time step 
\begin{equation}
\Delta t  = R_S \frac{ (\Delta x)^2 }{ 6 \widetilde D } ,
\end{equation}
where $0 < R_S < 1$ is a constant safety factor, here set to 0.9. 
Compared to the model as presented in earlier articles~\cite{Echebarria04, Echebarria10}, we use additional methods to enhance computational efficiency. 
First, the PF model uses a preconditioned phase field $\psi$, instead of using the classical phase field $\varphi$, through the nonlinear change of variable $\varphi\equiv\tanh(\psi/\sqrt{2})$. 
This change of variable makes the calculations stable for coarser numerical grids~\cite{Glasner01}. 
Second, we use the CUDA programing platform~\cite{Cuda} to perform simulations in parallel on Nvidia\textsuperscript{TM} GPU (Graphics Processing Unit) architectures. 
For multi-GPU calculations, we use direct peer-to-peer communication between GPUs~\cite{Cuda:multigpu, ShimokawabeEtAl11}. 
For the simulation of large domains (e.g. simulation in Fig.~3 in the Letter), we use up to eight GPUs simultaneously. 
Within the explicit Euler time stepping scheme, the random perturbation on the $\psi$ field is introduced as 
\begin{equation}
\psi^{t + \Delta t} = \psi^{t}+ \Delta t \partial_t \psi + F_{\psi} \sqrt{\Delta t}  \zeta ,
\end{equation}
where $F_{\psi}$ is a noise strength, and $\zeta$ is a random number, generated at each grid point and each time step with a uniform distribution, within the $[-0.5,+0.5]$ range.

\subsection{Crystal orientation in the numerical model.} 
\begin{figure}
\centering
\includegraphics[width = 136 mm]{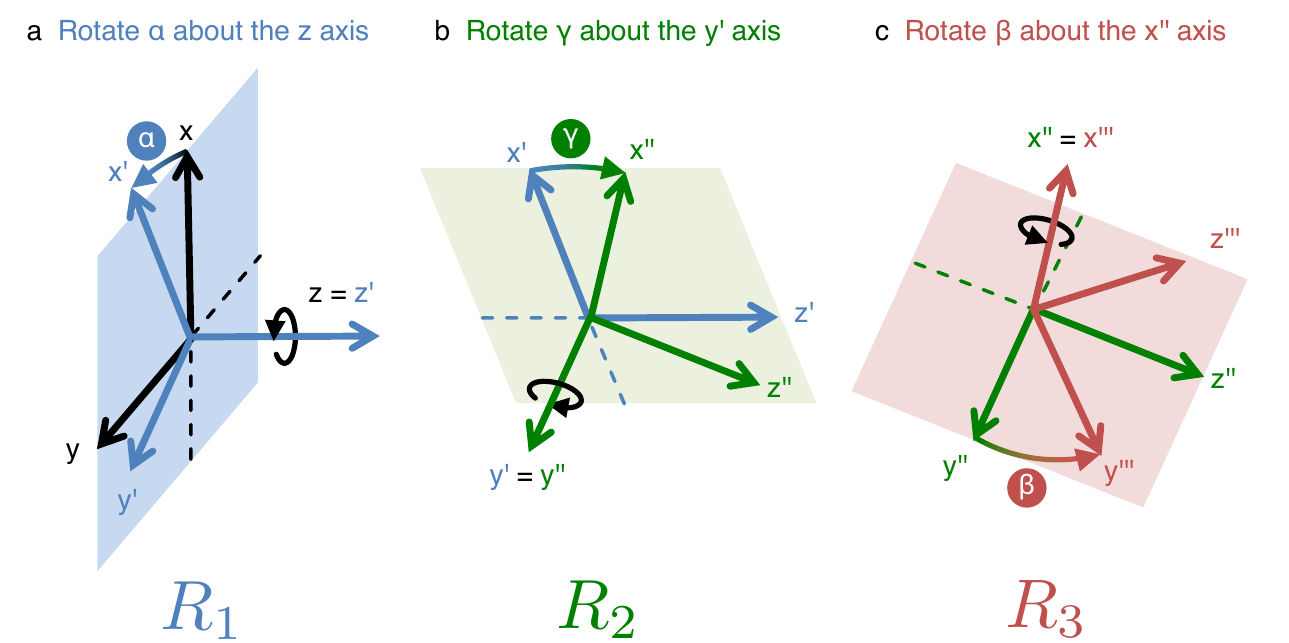}%
\caption{ 
\label{fig:rotation}
\rm
{\bf Rotated crystal axes $(x''',y''',z''')$ with respect to the fixed Cartesian axes $(x,y,z)$.}
The crystal orientation axes $(x''',y''',z''')$ are obtained by successively applying rotations: 
$R_1$ by an angle $\alpha$ about the $z$ axis, resulting in axes $(x',y',z')$ illustrated in {\suba}; 
$R_2$ by an angle $\gamma$ about the $y'$ axis, resulting in axes $(x'',y'',z'')$ illustrated in {\subb}; 
$R_3$ by an angle $\beta$ about the $x''$ axis, resulting in axes $(x''',y''',z''')$ illustrated in {\subc}.
Therefore, the red arrows in {\subc} indicate the rotated crystal axes. 
}
\end{figure}

$\, $ 
{\it Solid-liquid interface anisotropy.}
We use a standard form of the surface tension anisotropy with cubic symmetry given by~\cite{KarmaLeePlapp00}
\begin{equation} 
a_s(\mathbf{n}) = (1-3\epsilon_4)\left[ 1+\frac{4\varepsilon_4}{1-3\varepsilon_4}\left( n_x^4+n_y^4+n_z^4 \right) \right] ,
\end{equation}
where $\varepsilon_4$ is a measure of the anisotropy strength, and $\mathbf{n} =  (n_x, n_y, n_z)$ is the unit vector normal to the interface pointing towards the liquid, where $n_x$, $n_y$, and $n_z$ are components of $\mathbf{n}$ along the $x$, $y$, and $z$ axes, respectively.
To impose the three-dimensional (3D) orientation of a grain in the PF model, we use three orientation angles $\alpha$, $\gamma$, and $\beta$, following three successive steps: (1) rotate about the $z$-axis by an angle $\alpha$ (Supplementary Fig.~\ref{fig:rotation}{\suba}), (2) rotate about the $y'$-axis by an angle $\gamma$ (Supplementary Fig.~\ref{fig:rotation}{\subb}), and (3) rotate about the $x''$-axis by an angle $\beta$ (Supplementary Fig.~\ref{fig:rotation}{\subc}). 
Resulting crystal axes $(x''',y''',z''')$ are thus rotated by angles $(\alpha, \beta, \gamma)$ with respect to the fixed Cartesian axes $(x,y,z)$ as schematically illustrated in Supplementary Fig.~\ref{fig:rotation}. 
The $x'''$ direction represents the main growth direction of a grain.
This $x'''$ axis is used to identify the preferred growth direction using spherical angles $\theta$ and $\phi$, as explained later. 


{\it Implementation of the rotated anisotropy.}
The rotation with angles $(\alpha, \gamma, \beta)$ can be written in matrix forms as
\begin{equation}
\label{eqn:alpha}
\left(
\begin{array}{c}
x'   \\
y'  \\
z'
\end{array}
\right) 
=
R_1
\left(
\begin{array}{c}
x   \\ y  \\ z
\end{array}
\right) 
=
\left(
\begin{array}{ccc}
\cos\alpha  & \sin\alpha  & 0  \\
-\sin\alpha  & \cos\alpha & 0 \\
0                &        0          & 1
\end{array}
\right) 
\left(
\begin{array}{c}
x   \\ y  \\ z
\end{array}
\right) 
,
\end{equation}
\begin{equation}
\left(
\begin{array}{c}
x''   \\ y''  \\ z''
\end{array}
\right) 
=
R_2
\left(
\begin{array}{c}
x'  \\ y'  \\ z'
\end{array}
\right) 
=
\left(
\begin{array}{ccc}
\cos \gamma & 0 & \sin \gamma  \\
0                &       1          & 0 \\
- \sin \gamma  & 0 &   \cos \gamma 
\end{array}
\right) 
\left(
\begin{array}{c}
x' \\ y' \\ z'
\end{array}
\right) 
,
\end{equation}
and
\begin{equation}
\label{eqn:beta}
\left(
\begin{array}{c}
x'''  \\ y'''  \\ z'''
\end{array}
\right) 
=
R_3
\left(
\begin{array}{c}
x'' \\ y'' \\ z''
\end{array}
\right) 
=
\left(
\begin{array}{ccc}
1 &      0         &  0  \\
0 & \cos\beta & \sin\beta \\
0 & - \sin\beta & \cos\beta
\end{array}
\right) 
\left(
\begin{array}{c}
x'' \\ y'' \\ z''
\end{array}
\right) 
.
\end{equation}
The crystal axes $(x''', y''', z''')$ thus derive from the fixed Cartesian axes $(x, y, z)$ through
\begin{equation}
\left(
\begin{array}{c}
x''' \\ y''' \\ z'''
\end{array}
\right) 
=
\mathbf{R}
\left(
\begin{array}{c}
x  \\ y  \\ z
\end{array}
\right) 
,
\end{equation}
where the full rotation matrix is
\begin{equation}
\mathbf{R}
=
R_3 R_2 R_1
=
\left(
\begin{array}{ccc}
\label{eqn:rotarray}
r_{11} & r_{12} & r_{13} \\
r_{21} & r_{22} & r_{23} \\
r_{31} & r_{32} & r_{33}
\end{array}
\right) 
\end{equation}
with
\begin{align*}
r_{11} &= \cos\alpha \cos\gamma ,
\\
r_{12} &= \sin\alpha  \cos \gamma ,
\\
r_{13} &=  \sin \gamma ,
\\
r_{21} &= - \sin\alpha \cos\beta - \cos \alpha \sin\beta \sin\gamma ,
\\
r_{22} &= \cos\alpha \cos\beta - \sin \alpha \sin\beta \sin\gamma ,
\\
r_{23} &=  \sin\beta \cos\gamma ,
\\
r_{31} &= \sin\alpha \sin\beta - \cos \alpha \cos\beta \sin\gamma ,
\\
r_{32} &= - \cos \alpha \sin\beta - \sin \alpha \cos\beta \sin\gamma , 
\\
r_{33} &= \cos\beta \cos \gamma .
\end{align*}
Inversely, the fixed Cartesian coordinates ($x$, $y$, $z$) can be retrieved by
\begin{equation}
\left(
\begin{array}{c}
x   \\
y  \\
z
\end{array}
\right) 
=
\mathbf{R}^{-1}
\left(
\begin{array}{c}
x'''   \\
y'''  \\
z'''
\end{array}
\right) 
=
\left(
\begin{array}{ccc}
r_{11} & r_{21} & r_{31} \\
r_{12} & r_{22} & r_{32} \\
r_{13} & r_{23} & r_{33}
\end{array}
\right) 
\left(
\begin{array}{c}
x'''   \\
y''' \\
z'''
\end{array}
\right) 
.
\end{equation}

\noindent Thus, in Eq.~\eqref{eqn:psi}, the first derivatives of $\psi$ with respect to crystal axes $(x''',y''',z''')$ are
\begin{align}
\partial_{x'''} \psi &= \partial_{x'''} x \partial_x \psi +\partial_{x'''} y \partial_y \psi +\partial_{x'''} z \partial_z \psi ,
\\
\partial_{y'''} \psi &= \partial_{y'''} x \partial_x \psi +\partial_{y'''} y \partial_y \psi +\partial_{y'''} z \partial_z \psi ,
\\
\partial_{z'''} \psi &= \partial_{z'''} x \partial_x \psi +\partial_{z'''} y \partial_y \psi +\partial_{z'''} z \partial_z \psi ,
\end{align}
or in a matrix form
\begin{equation}
\label{eqn:firstderiv}
\left(
\begin{array}{c}
\partial_{x'''}    \\ \partial_{y'''}   \\ \partial_{z'''}
\end{array}
\right) 
\psi 
=
\left(
\begin{array}{ccc}
r_{11} & r_{12} & r_{13} \\
r_{21} & r_{22} & r_{23} \\
r_{31} & r_{32} & r_{33}
\end{array}
\right) 
\left(
\begin{array}{c}
\partial_{x}   \\ \partial_{y}  \\ \partial_{z}
\end{array}
\right) 
\psi .
\end{equation}

\noindent The second order derivative with respect to $x'''$ can be calculated as
\begin{align}
\partial_{x''' x'''} \psi &= r_{11}  \partial_x (\partial_{x'''} \psi)  + r_{12} \partial_y  (\partial_{x'''} \psi)  + r_{13} \partial_z  (\partial_{x'''} \psi)  .
\end{align}
Similarly, developed expressions of second derivatives are
\begin{align*}
\partial_{x''' x'''} \psi &= r_{11} ^2 \partial_{x x} \psi + r_{12}^2 \partial_{y y} \psi + r_{13}^2 \partial_{z z} \psi 
+ 2 ( r_{11} r_{12} \partial_{x y} \psi + r_{12} r_{13} \partial_{y z} \psi  + r_{11} r_{13} \partial_{z x} \psi  ) ,
\\
\partial_{y''' y'''} \psi  &= r_{21} ^2 \partial_{x x} \psi + r_{22}^2 \partial_{y y} \psi + r_{23}^2 \partial_{z z} \psi 
+ 2 ( r_{21} r_{22} \partial_{x y} \psi + r_{22} r_{23} \partial_{y z} \psi  + r_{21} r_{23} \partial_{z x} \psi  ) ,
\\
\partial_{z''' z'''} \psi  &= r_{31} ^2 \partial_{x x} \psi + r_{32}^2 \partial_{y y} \psi + r_{33}^2 \partial_{z z} \psi 
+ 2 ( r_{31} r_{32} \partial_{x y} \psi + r_{32} r_{33} \partial_{y z} \psi  + r_{31} r_{33} \partial_{z x} \psi  ) ,
\\
\partial_{x''' y'''} \psi  &= r_{11} ( r_{21} \partial_{x x} \psi + r_{22} \partial_{x y} \psi + r_{23} \partial_{x z} \psi)
+ r_{12} ( r_{21} \partial_{y x} \psi + r_{22} \partial_{y y} \psi+ r_{23} \partial_{y z} \psi)
\\
&+ r_{13} ( r_{21} \partial_{z x} \psi + r_{22} \partial_{z y} \psi+ r_{23} \partial_{z z} \psi) ,
\\
\partial_{y''' z'''} \psi  &= r_{21} ( r_{31} \partial_{x x} \psi + r_{32} \partial_{x y} \psi+ r_{33} \partial_{x z} \psi)
+ r_{22} ( r_{31} \partial_{y x} \psi + r_{32} \partial_{y y} \psi+ r_{33} \partial_{y z} \psi)
\\
&+ r_{23} ( r_{31} \partial_{z x} \psi + r_{32} \partial_{z y} \psi+ r_{33} \partial_{z z} \psi) ,
\\
\partial_{z''' x'''} \psi &= r_{31} ( r_{11} \partial_{x x} \psi + r_{12} \partial_{x y} \psi+ r_{13} \partial_{x z} \psi)
+ r_{32} ( r_{11} \partial_{y x} \psi + r_{12} \partial_{y y} \psi+ r_{13} \partial_{y z} \psi)
\\
&+ r_{33} ( r_{11} \partial_{z x} \psi + r_{12} \partial_{z y} \psi+ r_{13} \partial_{z z} \psi) .
\end{align*}
Here we only studied configurations with $\beta=0^\circ$ while exploring $\alpha$ and $\gamma$ orientation angles, such that coefficients of the rotation matrix may be simplified to $r_{11} = \cos\alpha \cos\gamma$, $r_{12} = \sin\alpha  \cos \gamma$, $r_{13} =  \sin \gamma$, $r_{21} = - \sin\alpha$, $r_{22} = \cos\alpha$, $r_{23} =  0$, $r_{31} = - \cos \alpha \sin\gamma$, $r_{32} = - \sin \alpha \sin\gamma$, and $r_{33} = \cos \gamma$.

{\it Bi-crystalline implementation.} 
We distinguish the two grains in the simulations by introducing an integer grain index field $p(x,y,z,t)$ ~\cite{TourretKarma15, Tourret17}. 
In the liquid far from the solid-liquid interface, the index value is initialized as $p_{i,j,k} = 0$ at a location of discrete spatial coordinates $(i,j,k)$ along the $x$, $y$, and $z$ axes, respectively. 
This value is updated as soon as $(1- \varphi_{i,j,k}^2 ) \ge 0.001$, i.e. when the solid-liquid interface approaches.
The new index value is estimated by using the summation of $p$ indices across the 27 neighbors 
\begin{align}
S_{i,j,k} = {\rm sign}\left\{ \sum_{ i-1}^{i+1} \, \sum_{ j-1}^{j+1} \, \sum_{k-1}^{k+1} \, p_{(i, j, k)}  \right\}, 
\end{align}
such that the index is then fixed to $p = - 1$ in one grain and $p = + 1$ in the other grain.
The current approach thus does not account for triple junctions and grain boundary motion in the solid state, which would anyway be inexistent due to the assumption of no solid-state diffusion.

{\it Conversion to experimentally measured spherical angles.}
When $\alpha$ and $\gamma$ have non-zero values, the main crystal orientation axis, i.e. the $x'''$ axis (red arrow in Fig.~4 in the Letter), differs from the $x$ axis of the temperature gradient. 
For convenience, we convert angles ($\alpha$, $\gamma$) at $\beta = 0^\circ$ introduced in previous subsections into spherical angles ($\theta$, $\phi$), directly extracted from the experimental measurements, and used throughout the discussion in the Letter.
From the components $(r_{11}, r_{12}, r_{13})$ of Eq.~\eqref{eqn:rotarray}, one can readily calculate spherical angles ($\theta$, $\phi$) using standard geometrical relationships, such as $\cos \theta = r_{11}$ and $\tan \phi = (r_{13}/r_{12})$. 
Supplementary Table~\ref{tab:Angles} summarizes ($\alpha$, $\gamma$) and calculated ($\theta$, $\phi$) angles used in PF simulations. 
In single-grain simulations, $\phi$ is the (counterclockwise) angle with respect to the $y+$ direction. 
In bi-crystalline simulations, the orientation $\phi_1$ of grain 1 (represented in blue) is measured with respect to the $y+$ direction (as in single-grain simulations), while the orientation $\phi_2$ of grain 2 (represented in red) is measured from the $y-$ direction, both of them measured counterclockwise, as illustrated in Fig.~4 in the Letter.
Note that we always set $\beta = 0^\circ$, which does not have a strong influence in the cellular regime under investigation here, but would matter in the dendritic regime when individual dendrites exhibit non-axisymmetrical features such as fins and sidebranches~\cite{KarmaLeePlapp00}.



\subsection{Simulation parameters.}

We use material parameters for a SCN-0.24wt\%~camphor alloy identified and discussed in previous studies~\cite{TourretEtAl15,PeredaEtAl17, Mota15, Mota16, Mota17}. 
These parameters are listed in Supplementary Table~\ref{tab:params}. 
%
%
PF simulations exclusively focus on one set of experimental parameters, namely with $V = 1.5$~{\utv} and $G = 19$~{\utG}. 
The corresponding numerical parameters are also listed in Supplementary Table~\ref{tab:params}. 
Simulations use a random noise strength $F_\psi = 0.01$, except for simulations in Supplementary Fig.~\ref{fig:dirftangles} that are performed with $F_\psi = 0$. 

\subsection{Simulation configurations.}
\begin{figure}
\centering
\includegraphics[width = 89 mm]{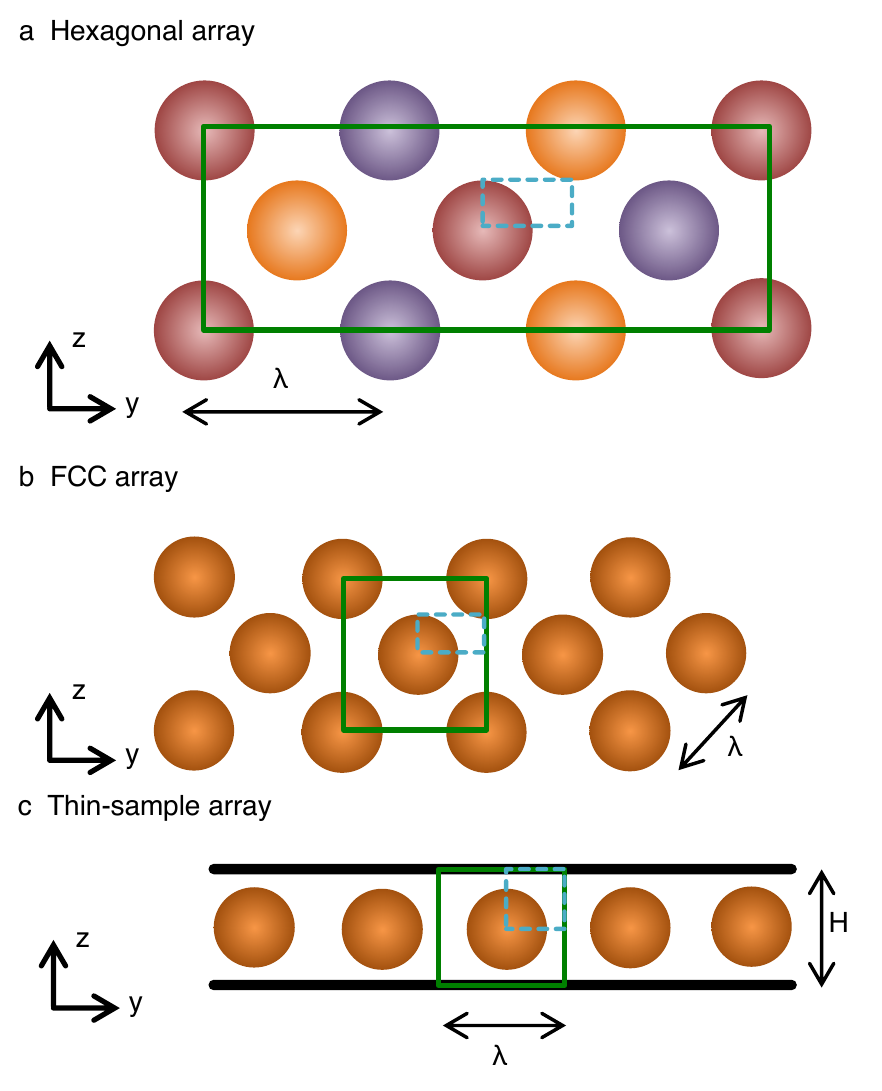}
\caption{ 
\label{fig:schemearray}
\rm
{\bf Schematics of different 3D pattern symmetries.}
We considered three symmetric patterns of hexagonal (a), FCC-like (face centered cubic) (b), and thin-sample (c) arrays.
The thicker black lines in c indicate the sample walls. 
We performed PF simulations with a domain (green square and cyan square) using symmetry conditions (details in Supplementary Methods).
}
\end{figure}

$\, $ 
{\it Drift dynamics of a single grain.}
The relationship between the crystal orientation of a grain $\theta$ and its actual growth direction $\theta_d$ was initially derived from confined thin sample experiments~\cite{Akamatsu97, Deschamps08} and afterwards validated both 2D and confined 3D phase-field simulations~\cite{TourretKarma15, GhmadhEtAl14,LiWang12,SongEtAl18:PRMater}.
Here, we further validated that this relationship still stands for different 3D patterns.
To do so, we generated microstructure arrays growing at steady state at $V = 1.5$~{\utv} and $G = 19$~{\utG} with different pattern symmetry, namely hexagonal (Supplementary Fig.~\ref{fig:schemearray}{\suba}) and FCC-like (faced centered cubic, Supplementary Fig.~\ref{fig:schemearray}{\subb}), as well as a thin-sample confined-3D geometry (Supplementary Fig.~\ref{fig:schemearray}{\subc}).
Using the resulting arrays of imposed primary spacing $\lambda$ generated with a well-oriented grain with $\theta=\phi=0^\circ$ as initial conditions, we then restarted the simulations with different $(\theta,\phi)$ orientations in order to assess the dependence of $\theta_d$ upon $\theta$, $\phi$, and $\lambda$.
The procedure is, to a large extent similar to that presented in previous papers~\cite{TourretEtAl15}.
First, we simulated the steady state growth of one quarter of a cell using symmetry conditions appropriate for different array structures, illustrated as cyan dashed boxes in Supplementary Fig.~\ref{fig:schemearray}. 
These simulations were performed for a grain oriented along the temperature gradient direction, i.e. with $\theta=\phi=0^\circ$.
For a hexagonal array (Supplementary Fig.~\ref{fig:schemearray}{\suba}), we used a domain size ratio of $L_z/L_y = \sqrt{3}/2$, with $L_y=\lambda/2$~\cite{Gurevich10,TourretEtAl15}.
For an FCC-like array (Supplementary Fig.~\ref{fig:schemearray}{\subb}), we used a domain size ratio $L_z/L_y = 1/2$, with $L_y = \lambda \sqrt{2}$.
These simulations were performed with no-flux boundary conditions along the $y$ boundaries and the $z=0$ boundary and anti-symmetry condition along the $z = L_z$ boundary (see Refs.~\cite{BergeonEtAl13,TourretEtAl15}). 
For a thin-sample geometry (Supplementary Fig.~\ref{fig:schemearray}{\subc}) we used $L_y = \lambda/2$ and a fixed $L_z = H/2$ and no-flux boundary conditions along all boundaries except for the sample walls (thick black lines), where we set a simplified geometrical wetting condition by imposing a non-normal angle between the solid-liquid interface and the wall (see Refs.~\cite{TourretEtAl15, Tourret17, GhmadhEtAl14}):
\begin{equation}
\label{eq:wetting}
\left. \frac{\partial \psi}{\partial z} \right |_{z = 0} = -  \left. \frac{\partial \psi}{\partial z} \right |_{z = H} = +1 .
\end{equation}
Then, after obtaining one quarter of a steady state cell for one given $\lambda$, we used the resulting fields as initial conditions of a new simulation for a slightly different $\lambda$. 
In order to reduce (increase) $\lambda$ while keeping numerical parameters $\Delta x$ and $W$ unchanged for the sake of a consistent numerical accuracy, we restarted simulations with fewer (additional) numerical grid points.
Hence, each new simulation was started from $\psi$ and $U$ fields that were either stretched or squeezed and thus bilinearly interpolated from the previous simulation for a grid with a slightly different number of points in $y$ and $z$ dimensions, while the $x$ dimension remained unchanged throughout.
We increased $L_y$ from $70$~{\utmicron} to $160$~{\utmicron} for a hexagonal array, and from $70$~{\utmicron} to $211$~{\utmicron} for an FCC array, both with steps of $13$~{\utmicron}.
For the thin-sample array, we increased $L_y$ from $67$~{\utmicron} to $163$~{\utmicron} with steps of $6$~{\utmicron}, for a given sample thickness $H=198$~{\utmicron}.
%
%
Finally, we duplicated fields obtained for a steady-state quarter of cell in order to build entire stable arrays, illustrated as green thin boxes in Supplementary Fig.~\ref{fig:schemearray}.
Then, we imposed crystal angles ($\alpha$, $\gamma$), or ($\theta$, $\phi$), in order to investigate the drift dynamics of cells in an entire array. 
Periodic boundary conditions were imposed along the $y$ and $z$ axes for hexagonal and FCC arrays. 
The simulations of the thin-sample array has periodic conditions along the $y$ boundaries and an imposed wetting angle along the sample walls at the $z$ boundaries, following Eq.~\eqref{eq:wetting}. 
We use these simulations to study the selection of the cell growth direction, i.e. the pattern drift direction and velocity linked to $\theta_d$, as discussed later. 

{\it Initial conditions for a spatially-extended bi-crystal.}
To perform bi-crystal simulations such as in Fig.~3 and Fig.~4 of the Letter, we used a similar method as described in Ref.~\cite{Tourret17}. 
The boundary conditions were symmetric (i.e. no-flux) for the $x$ and $y$ boundaries and periodic for the $z$ boundaries.
First, we run a preliminary simulation with both the left (blue) and right (red) grains well-oriented, i.e. with $\theta = \phi=0^\circ$.
A planar two-sided no-flux inner boundary, normal to the $y$ axis, is located between the two grains at $y=y_b$ in order to prevent premature invasion of grains during transient microstructure growth after the destabilization of the planar front~\cite{Tourret17}.
The inner boundary is fixed within the sample frame was set with a length of $2000$~{\utmicron} in the $x$ direction.
The simulation was then performed slightly longer after the inner boundary disappearance in order to allow the pattern to stabilize and avoid transient conditions relative to the removal of this inner wall. 
At the end of the simulation, i.e. at $t = 3000~{\rm s}$, the GB was hence relatively straight and normal to the $y$ axis and located around the inner boundary position $y_b$. 
For the simulation reproducing experimentally measured grain orientations in Fig.~3 of the Letter, the domain size is $L_x \times L_y \times L_z = 2295 \times 1912 \times 633~$\textmu m$^3$, and the initial GB is set at $y_b = 478$~{\utmicron}.
For the mapping of GB stability as a function of bi-crystal configurations $(\phi_1,\phi_2)$ in Fig.~4{\subb} of the main text, the domain size is with $L_y = 1272$~{\utmicron} with similar $L_x$ and $L_z$ as previously.
The initial GB location was then set to $y_b = 636$~{\utmicron}, i.e. in the middle of $L_y$.
While all simulations presented in Fig.~4{\subb} of the Letter were performed with $y_b=L_y/2$, additional simulations discussed later in this document (Supplementary Fig.~\ref{fig:GCTheta}{\suba}-{\subc}) were performed with $y_b = 3L_y/4=959$~{\utmicron}. 
We used the resulting $\psi$, $U$, and $p$ fields as initial conditions of the bi-crystal simulations in the Letter, only changing the grain orientations to the desired values, i.e. $(\theta_1,\phi_1)=(6^\circ, -56^\circ)$ and $(\theta_2,\phi_2)=(3^\circ, 84^\circ)$ in Fig.~3, and $\theta_1=\theta_2=5^\circ$ with various $(\phi_1,\phi_2)$ combinations in Fig.~4{\subb}.

{\it Solitary cell in a host hexagonal grain.}
In order to produce a SC within a foreign hexagonal array, we first built an extended 24-cell array, hence four times larger than the green-thick box in Supplementary Fig.~\ref{fig:schemearray}{\bf a}, using a similar procedure as described in the previous subsection.
From the resulting $\psi$ field obtained for a single grain with $\theta=\phi=0^\circ$, we used a separate bespoke C++ code to read the $\psi$ field and isolate a single cell, in order to build a grain index field with a single (blue) cell with $p = +1$ and the remaining 23 (red) cells with $p = -1$.
Using these fields, we let the new initially imposed index field relax to a steady state by performing a simulation over $t = 3600$~s, thus obtaining a steady-state array structure with a single SC, still with $\theta=\phi=0^\circ$ for both grains (images in the first row of Supplementary Fig.~\ref{fig:scs}). 
From there, we restarted the simulation changing the orientation of the blue cell to $\theta_1=5^\circ$ and different $\phi_1$ while keeping $\theta_2=\phi_2=0^\circ$ in the host (red) array.

\section*{Supplementary Notes}

\subsection{Supplementary Note 1: Growth orientation selection in 3D arrays.}

\begin{figure}
\centering
\includegraphics[width = 89 mm]{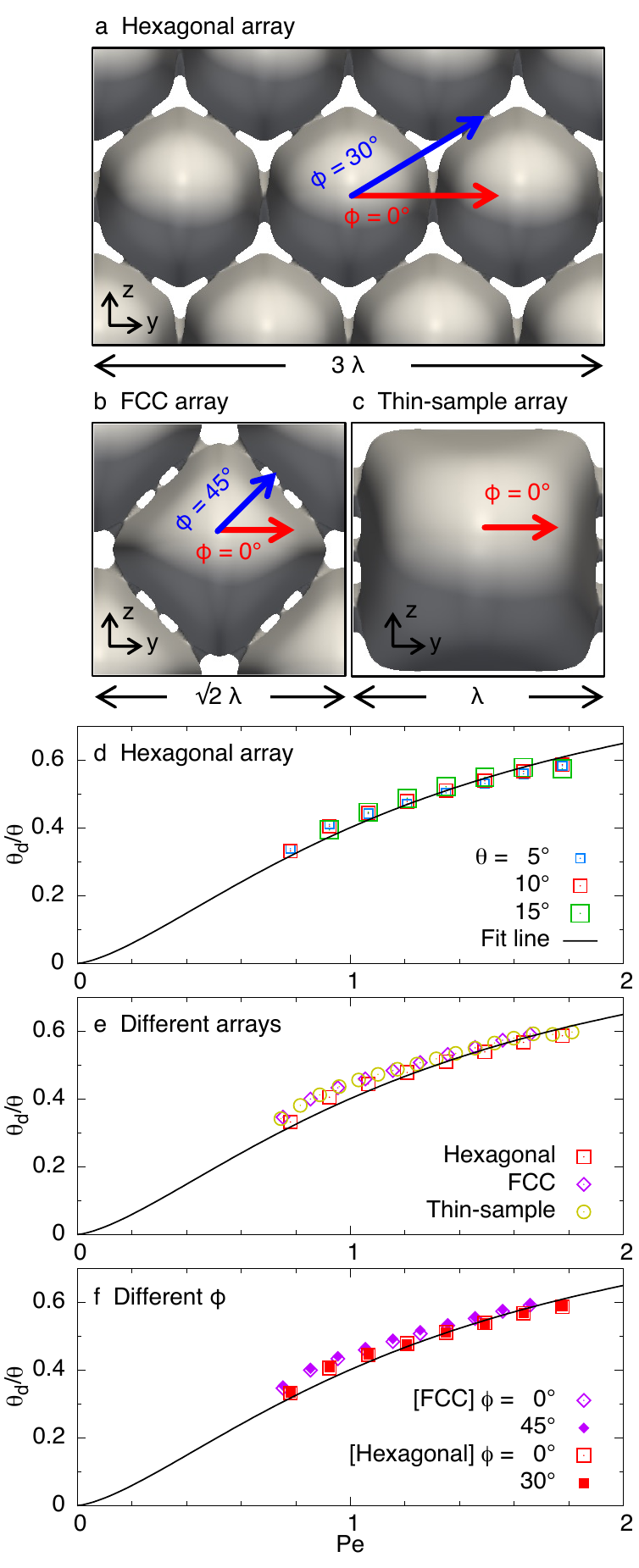}
\caption{ 
\label{fig:dirftangles}
\rm
{\bf Lateral drift dynamics of different patterns.}
Typical microstructures for different imposed array symmetries, namely hexagonal ({\suba}), FCC-like ({\subb}), and confined thin sample ({\subc}). 
Drift directions $\theta_d$ relative to the crystal orientation $\theta$ as a function of the P\'eclet number $\Pe = \lambda V/D$ for different $\theta$ ({\subd}), array symmetries ({\sube}), and projected drifting directions $\phi$ ({\subf}).
Thin-sample configurations are with a fixed sample thickness $H = 198$~{\utmicron}.
Simulations in {\subd}-{\sube} are with $\phi = 0^\circ$ (open symbols).
Simulations in {\sube}-{\subf} are with $\theta = 10^\circ$.
Two simulations in {\subf} are with $\phi = 30^\circ$ for a hexagonal array and $\phi = 45^\circ$ for a FCC array (closed symbols).
The black line in {\subd}-{\subf} shows the best fit of PF results with a hexagonal array for ($\theta$, $\phi$) = ($10^\circ$, $0^\circ$) (red open squares) to Eq.~\eqref{eqn:driftangles}, i.e. with $f = 0.67$ and $g = 1.47$.
}
\end{figure}

From the results of PF simulations of a single tilted grain, we show that the scaling laws for the dependence of the growth angle $\theta_d$ upon the crystal angle $\theta$ derived from confined thin sample experiments~\cite{Akamatsu97, Deschamps08} and validated with similarly confined phase-field simulations~\cite{TourretKarma15, GhmadhEtAl14,LiWang12, SongEtAl18:PRMater}, still hold for three-dimensional spatially-extended patterns~\cite{MotaEtAl21}.
The scaling law relating the ratio $\theta_d/\theta$ to the array P\'eclet number $\Pe = \lambda V/D$ writes
\begin{equation}
\label{eqn:driftangles}
\frac{\theta_d }{ \theta } = 1 - \frac{1}{1 + f \, \Pe^g } ,
\end{equation}
where here we consider $f$ and $g$ as constants. 
We considered the case of a hexagonal array (Supplementary Fig.~\ref{fig:dirftangles}{\suba}), a FCC-like array (Supplementary Fig.~\ref{fig:dirftangles}{\subb}) and a confined thin-sample array (Supplementary Fig.~\ref{fig:dirftangles}{\subc}). 
Red and blue arrows in Supplementary Fig.~\ref{fig:dirftangles}{\suba}-{\subc} illustrate drift directions $\phi$ that were considered. 
%
%
In Supplementary Figs.~\ref{fig:dirftangles}{d}-{f}, we plot $\theta_d/\theta$ as a function of {\Pe} for different $\theta$ using a hexagonal array ({\subd}), for different array symmetry with $\theta=10^\circ$ ({\sube}), and for different drift angles $\phi$ in hexagonal and FCC arrays ({\subf}).
All data points in Supplementary Figs.~\ref{fig:dirftangles}{d}-{f} collapse onto one master curve defined by Eq.~\eqref{eqn:driftangles} and shown as the black solid line, with values of $f = 0.67$ and $g = 1.47$ obtained from the least-square fit of the PF results with a hexagonal array for ($\theta$, $\phi$) = ($10^\circ$, $0^\circ$).
These results confirm that Eq.~\eqref{eqn:driftangles} still stands for 3D configurations, and hence provide a direct way to correlate the crystal orientation angle $\theta$ input to our PF simulations with the growth direction angle $\theta_d$ estimated from drifting velocities measured in the experiments.

\subsection{Supplementary Note 2: Grain boundary stability.}

\begin{figure}
\centering
\includegraphics[width = 89 mm]{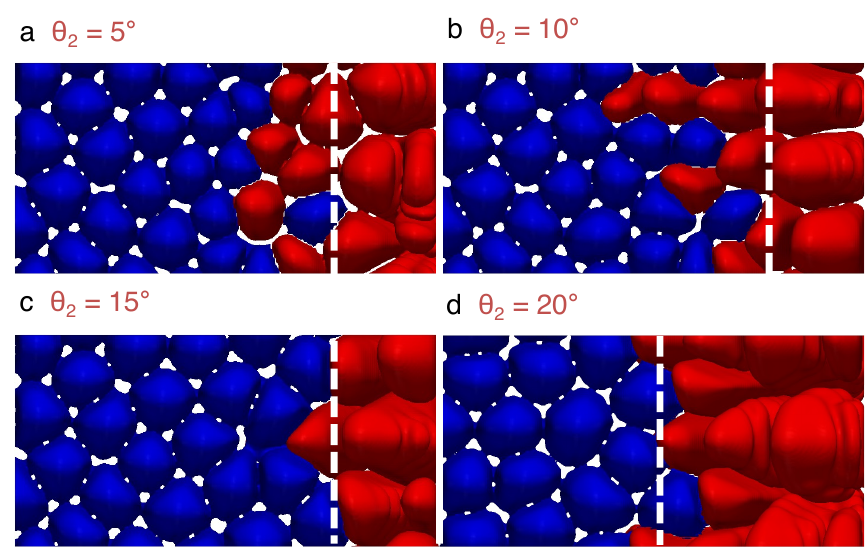}
\caption{ 
\label{fig:GCTheta}
\rm
{\bf GB morphology and stability with respect to the lateral drift of a grain.}
Images show the microstructures after $3$~{\uth} of growth using $\theta_2 = 5^\circ$ ({\suba}), $10^\circ$ ({\subb}), $15^\circ$ ({\subc}), and $20^\circ$ ({\subd}) for the right (red) grain.
In all cases $\theta_1 = \phi_1 = \phi_2 = 0^\circ$.  
White dashed lines indicate the initial GB position.
}
\end{figure}

$\, $ 
 {\it Influence of $\theta$.}
First, we investigated the influence of $\theta$ on the GB stability. 
Restarting from a simulation with a straight GB located between two well-oriented grains, we imposed different crystal orientations $\theta_2 > 0^\circ$ with $\phi_2 = 0^\circ$ for one grain while keeping $\theta_1 = \phi_1 = 0^\circ$ for the other grain. 
Supplementary Fig.~\ref{fig:GCTheta} shows the resulting microstructures after $3$~{\uth} for $\theta_2 = 5$, $10$, $15$, and $20^\circ$, as compared to the initial GB location (white dashed line).
When $\theta_2 \le 10^\circ$, red cells penetrate into the blue well-oriented grain (Supplementary Fig.~\ref{fig:GCTheta}{\suba}-{\subb}). 
This leads to the morphological destabilization of the initially straight GB (white dashed lines). 
On the other hand, when $\theta_2 \ge 15^\circ$, the GB remains relatively straight and close to its initial position (Supplementary Fig.~\ref{fig:GCTheta}{\subc}-{\subd}). 
This is consistent with the results of previous 2D and confined-3D thin sample simulations~\cite{Tourret17}, which showed that a converging GB has a high mobility only in the vicinity of a symmetrical configuration, e.g. for $\theta_1+\theta_2= 0^\circ$ and $\phi_1=\phi_2=0^\circ$.
In the case of Supplementary Fig.~\ref{fig:GCTheta}, since $\theta_1=0^\circ$, the converging GB only exhibits a high mobility in the vicinity of $\theta_2\approx0^\circ$, namely here for $\theta_2 \le 10^\circ$.
This is consistent with the experiment in Fig.~1{\subc} of the Letter, in which the left and right grains respectively have $\theta = 6$ and $3^\circ$. 
It is also worth noting that the penetration of a grain is expected to depend upon the primary spacing of the arrays~\cite{HuEtAl18,TakakiEtAl16:ISIJ}, which is $\lambda = 189$~{\utmicron} in these simulations. 


{\it Influence of $\phi$.}
%
As explained above, with the current parameters the (inter-)penetration of grain can only be observed for $\theta_2 \le 10^\circ$.
Hence, we studied the influence of $\phi_1$ and $\phi_2$ in bi-crystalline simulations with $\theta_1=\theta_2=5^\circ$ (Fig.~4{\subb} of the Letter).
Results of the simulations with a GB initially located near the middle of the simulation domain $L_y$ (white dashed line) after $3$~{\uth} of growth are illustrated in Supplementary Fig.~\ref{fig:GC3D}. 
%
%
The orientation of the blue grain is mapped from $\phi_1 = 0^\circ$ (left column) to $75^\circ$ (right column). 
The orientation of the red grain is mapped from $\phi_2 = -90^\circ$ (bottom row) to $+90^\circ$ (top row). 
These configurations are sufficient to reconstruct the entire $(\phi_1,\phi_2)$ space by symmetry and produce Fig.~4{\subb} of the Letter.
When one grain has $\phi\approx0^\circ$, i.e. when its drift direction is directly pointing towards the neighbor grain, groups of cells within the $\phi\approx0^\circ$ grain systematically penetrate the neighbor grain.
This can be seen for instance with $\phi_1 = 0^\circ$ in the leftmost column of Supplementary Fig.~\ref{fig:GC3D}, where the left (blue) grain systematically invades the right (red) grain.
Similarly, within the same first column, the right (red) grain may also penetrate the left (blue) grain, but only when its $\phi_2$ angle is also low enough, i.e. here for $\phi_2 \le 60^\circ$.
On the other hand, when $\phi_2 \ge 75^\circ$ for either of the grains, the penetration process remains limited.
Subsequently, the boundaries that remain the most stable and straight are for cases with both high $\phi_1$ and high $\phi_2$, i.e. in the top-right and bottom-right corners of Supplementary Fig.~\ref{fig:GC3D}.
For cases with $\phi_2 = \pm 90^\circ$ and $\phi_1 = 90^\circ$, which are not shown in Supplementary Fig.~\ref{fig:GC3D}, the GB remains stable and straight.


\subsection{Supplementary Note 3: Solitary cells.}
$\, $ 
{\it SC emergence.}
In Supplementary Fig.~\ref{fig:GC3D}, yellow circles identify the emergence of a SC or that of an isolated group of cells drifting inside the neighbor grain.
The observed region for possible emergence of a SC is represented with the thick green line, namely (1) for $\phi_1+\phi_2<90^\circ$ and (2) for both $|\phi_1|<75^\circ$ and $|\phi_2|<75^\circ$.

{\it Solitary cell behavior.}
We explored the behavior of an isolated SC as a function of the host array spacing $\lambda$ and of the SC drift direction $\phi_1$. 
In order to investigate these SC behaviors, we considered a SC within the host hexagonal array. 
The initial arrays had spacings $\lambda = 122$, 128, 134, 141, 147, 160, 192, 224, and 256~{\utmicron}, and cells in the host grain were favorably oriented (i.e. $\phi_2=\theta_2=0^\circ$). 
Then, we varied drift directions of one SC from $\phi_1= 0^\circ$ (close-packed direction) to $30^\circ$ at a fixed $\theta_1 = 5^\circ$. 
These simulations were carried out for at least $3$~{\uth} of growth --- some of them for up to $9$~{\uth} when necessary to unambiguously classify the behavior of the SC. 
Supplementary Fig.~\ref{fig:scs} illustrates three different behaviors for four different configurations (columns), starting from $t = 0$~{\uth} (top row) to later times (bottom rows). 
The initial array spacing is fixed at $\lambda = 141$~{\utmicron} for Supplementary Fig.~\ref{fig:scs}{\suba}-{\subc}. 
These results illustrate that the SC behavior depends upon its drift direction $\phi_1$ (white arrows in Supplementary Fig.~\ref{fig:scs}). 
For a higher $\phi_1 = 30^\circ$ ({\suba}), a SC drifts laterally until it is eliminated at $4.4$~{\uth}. Otherwise, a SC for a low $\phi_1 = 0^\circ$ ({\subb}) and $15^\circ$ ({\subc}) may progress as it eliminates neighbor cells (cross symbols). 
Even if a SC has a higher angle $\phi_1 = 30^\circ$, it may survive indefinitely when the host array has a larger spacing, e.g. for $\lambda = 192$~{\utmicron} ({\subd}).
%
%
In Supplementary Figs.~\ref{fig:scs}{\subb}-{\subc}, one host grain cell to be eliminated is identified with cross symbols on the center row.
This elimination may either affect a directly adjacent host cell if the drifting direction of the SC is closely directed toward the host cell, e.g. for $\phi_1 = 0^\circ$ ({\subb}), or the SC may first squeeze its way between initially adjacent cells before eliminating the host cell with the location closest to the SC path, e.g. for $\phi_1 = 15^\circ$ ({\subc}).
%
%
The resulting $(\phi_1,\lambda)$ map in Supplementary Fig.~\ref{fig:scsummary} shows three distinct behaviors. 
If an array has a spacing lower than the low spacing stability limit $\lambda_{min}$, a cell in the array is eliminated and the array spacing becomes higher~\cite{HuntLu96, Echebarria10, Clarke17, SongEtAl18:ActaMat}.  
We independently performed simulations to measure the limit of a hexagonal array, which yields $\lambda_{min} = 112$~{\utmicron} (gray dashed line). 
For SC simulations, when the initial spacing is closest to $\lambda_{min}$ (gray region), the SC is systematically eliminated as it tries to drift laterally within the array. 
On the other hand, when $\lambda$ is high enough (green region), the SC can move within the array and progresses as an accommodating SC by squeezing itself between red cells, without any elimination event. 
We expect that the SC would remain as an accommodating SC up to the higher spacing limit $\lambda_{max} = 342$~{\utmicron} (black dashed line). An array with $\lambda > \lambda_{max}$ would thus exhibit a sidebranching instability~\cite{HuntLu96, Echebarria10, Clarke17, SongEtAl18:ActaMat}. 
Within a narrow range of intermediate spacings (orange region), the SC progresses as a dominant SC by eliminating its first neighbor within the host array.
The dominant SC regime is the most common in experiments, while in simulations the SC often proceeds as an accommodating SC.
We attribute this discrepancy between experiments and simulations to potential uncertainties in material properties and thermal conditions, which are not directly measured during the experiments. For computational efficiency, present calculations consider a frozen temperature approximation (i.e. a one-dimensional temperature field with constant $G$ and $V$) and do not include the effect of latent heat release during solidification. Yet, previous studies have shown that relaxing these approximations may lead to a modest increase of the dynamically selected primary spacing~\cite{Mota15, SongEtAl18:ActaMat}. 
The conclusions drawn here regarding a SC behavior are valid for the specific case of a perfect hexagonal array and a perfectly oriented host grain.
It is expected that, as the topology of the array becomes more complex, for instance due to successive elimination and/or branching events, or when the host grain is also tilted, the classification of the SC behavior becomes more complex too. 


\break

\pagebreak

\section*{Supplementary Figures}

\pagebreak

\pagebreak

\pagebreak

\pagebreak

\pagebreak

\pagebreak
\begin{figure}
\centering
\includegraphics[width = 183 mm]{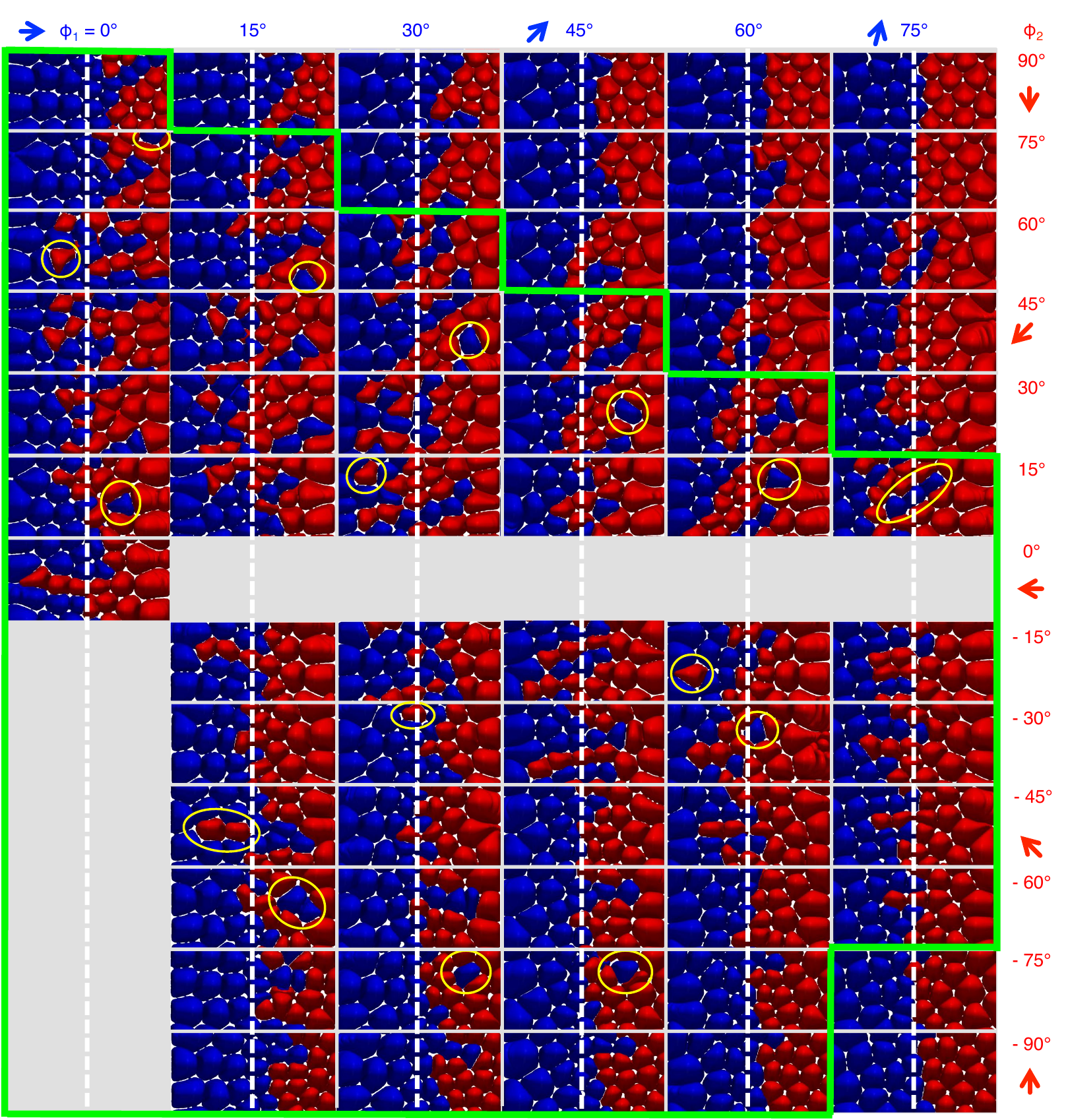}
\caption{ 
\label{fig:GC3D}
\rm
{\bf Bi-crystalline microstructures for different pattern drift angles ($\phi_1$, $\phi_2$).}
Images show simulated bi-crystalline microstructures after $3$~{\uth} of growth as a function of crystal orientation angles $\phi_1$ for the blue grain and $\phi_2$ for the red grain. 
The GB is initially located at the center of the domain (white dashed lines).
All simulations are with $\theta_1 = \theta_2 = 5^\circ$.
Solitary cells or groups are outlined in yellow.
The green solid line delimits the region of observed emergence of SCs.
}
\end{figure}

\pagebreak
\begin{figure}
\centering
\includegraphics[width = 183 mm]{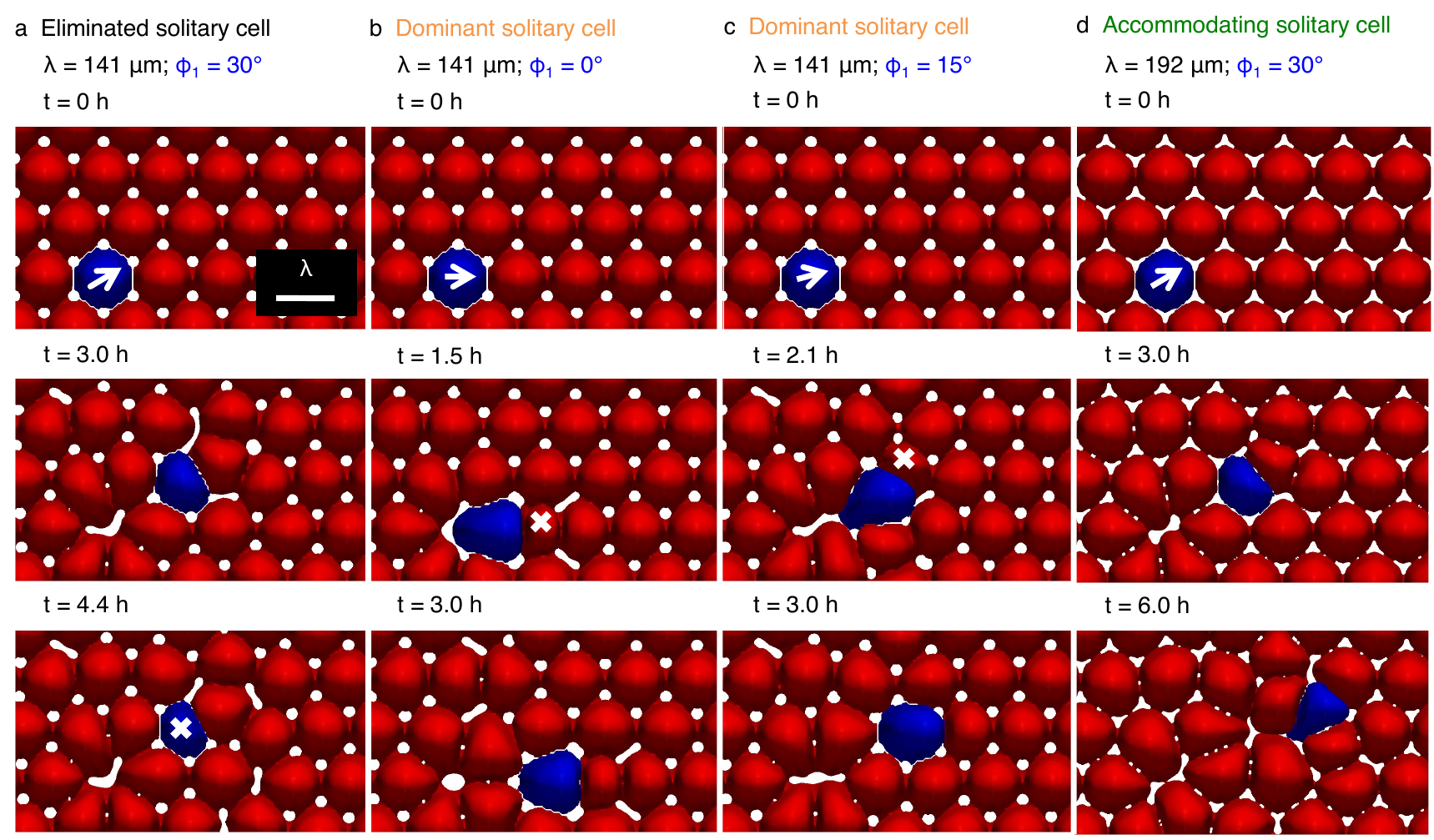}
\caption{ 
\label{fig:scs}
\rm
{\bf Behaviors of SCs with respect to their drift direction $\phi_1$ and to the host grain spacing $\lambda$.}
Images show simulated evolution of a (blue) SC within a (red) host array as time elapses (top to bottom row).
These simulations result in either: the elimination of the SC ({\suba}), a dominant SC that eliminates cells within the host array ({\subb}-{\subc}), or an accommodating SC that squeezes between the host array as it progresses without any elimination event ({\subd}).
Red hexagonal arrays are with $\lambda = 141$~{\utmicron} ({\suba}-{\subc}) and 192~{\utmicron} ({\subd}).
Drifting directions of SCs (white arrows on top-row images) are $\phi_1 = 0^\circ$ ({\subb}), $15^\circ$ ({\subc}), and $30^\circ$ ({\suba}, {\subd})%
, while $\theta_1=5^\circ$ and $\theta_2=\phi_2=0^\circ$. 
White crosses mark the eliminated cells shortly before their elimination.
}
\end{figure}

\pagebreak
\begin{figure}
\centering
\includegraphics[width = 89 mm]{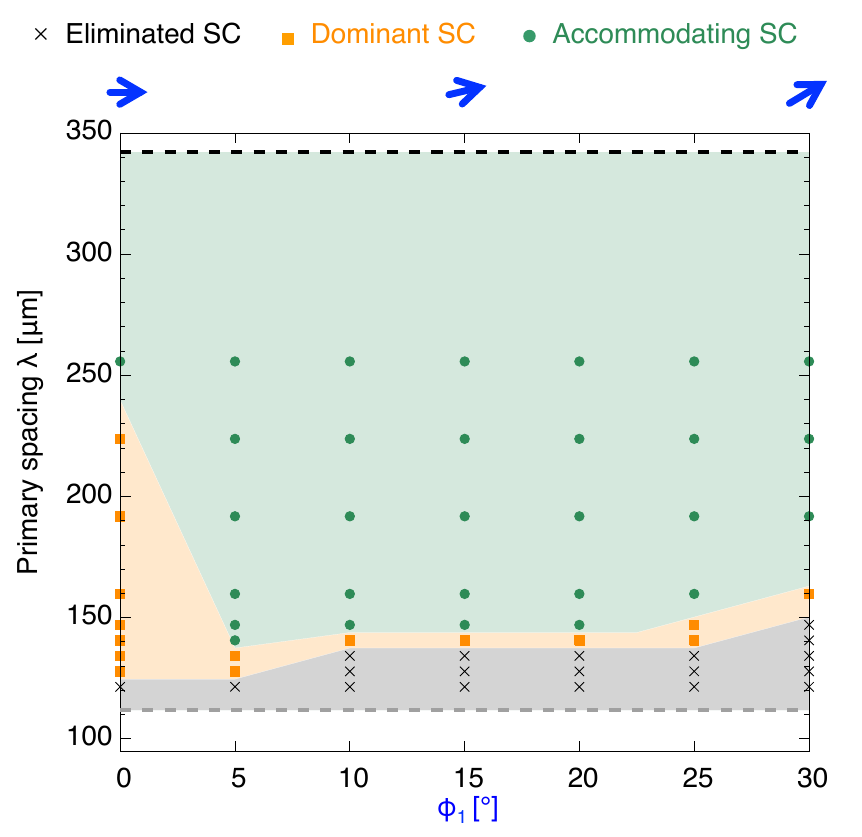}
\caption{ 
\label{fig:scsummary}
\rm
{\bf SC behaviors within a hexagonal host array.}
Using PF simulations, we identify three distinct behaviors (different colors) of SCs within a well-oriented ($\theta_2 = \phi_2 = 0^\circ$) hexagonal array. 
SCs are set with different $\phi_1$ for a fixed $\theta_1 = 5^\circ$.
Blue arrows above the graph illustrate the drift direction $\phi_1$. 
Simulations result in either: SC elimination (thin black crosses), or SC surviving with (orange squares) and without (green circles) elimination of cells in the host matrix. 
Horizontal gray and black dashed lines respectively mark the minimum $\lambda_{min}$ and maximum $\lambda_{max}$ stable spacing limits identified for a perfect hexagonal well-oriented array.
}
\end{figure}

\break 

\pagebreak

\section*{Supplementary Tables}

\begin{table}
\rm
\caption{
Material and numerical parameters for a SCN-0.24wt\%~camphor alloy at $V = 1.5$~{\utv} and $G = 19$~{\utG}. 
}
\begin{center}
\begin{tabular}{ l c c c  }
Parameter & Symbols & Value & Unit \\
\hline
\hline 
Liquidus slope & $m$ & -1.365 & K/wt\% \\
Diffusivity & $D$ & 270 & \textmu m$^2$/s \\
Gibbs-Thomson coefficient & $\Gamma$ &  64.78 & K {\utmicron} \\
Partition coefficient & $k$ & 0.07 &  \\
Anisotropy strength & $\varepsilon_4$ & 0.011 & \\
\hline
Interface thickness & $W$ & 179 & $d_0$ \\
Grid spacing & $\Delta x$ & 1.2 & $W$ \\
& & 3.2 & {\utmicron} \\
Time step & $\Delta t$ & 0.00567 & $s$ \\
Noise strength & $F_\psi$ & 0.01 & \\
\hline
\end{tabular}
\end{center}
\label{tab:params}
\end{table}

\break
\def\arraystretch{0.8}
\begin{table}
\rm
\caption{
Crystal orientations used in PF simulations. We imposed angles $\alpha$ and $\gamma$ into the PF model. These values are converted to spherical angles $\theta$ and $\phi$ (Supplementary Figs.~3 and Fig.~4a of the Letter).
}
\begin{center}
\begin{tabular}{ c c c c  }
~ $\alpha~[^\circ]$ ~ & ~ $\gamma~[^\circ]$ ~ & ~  $\theta~[^\circ]$ ~ & ~  $\phi~[^\circ]$ ~  \\
\hline
\hline 
0.0 & 0.0 & 0.0 & 0.0 \\
5.0 & 0.0 & 5.0 & 0.0 \\
4.9 & 0.4 & 4.9 & 4.7 \\
4.9 & 0.9 & 5.0 & 10.4 \\
4.9 & 1.3 & 5.1 & 14.9 \\
4.7 & 1.7 & 5.0 & 19.9 \\
4.5 & 2.1 & 5.0 & 25.0 \\
4.5 & 2.6 & 5.2 & 30.1 \\
3.6 & 3.6 & 5.1 & 45.1 \\
2.6 & 4.5 & 5.2 & 60.0 \\
1.3 & 4.9 & 5.1 & 75.2 \\
0.0 & 5.0 & 5.0 & 90.0 \\
\hline
10.0 & 0.0 & 10.0 & 0.0 \\
8.7   & 5.0 & 10.0 & 30.0 \\
7.1   & 7.1 & 10.0 & 45.2 \\
\hline
15.0 & 0.0 & 15.0 & 0.0 \\
\hline
0.3 & 2.8 & 2.8 & 83.9 \\
3.5 & 5.1 & 6.2 & 55.6 \\
\hline
\end{tabular}
\end{center}
\label{tab:Angles}
\end{table}


\pagebreak
\section*{Supplementary References}

